\documentclass[twocolumn]{autart}
\usepackage{graphicx}
\usepackage[T1]{fontenc}
\usepackage[applemac]{inputenc}
\makeatletter
\let\l@ENGLISH\l@english
\makeatother
\usepackage[english]{babel}
\usepackage[babel]{csquotes}
\usepackage{lipsum}
\usepackage{url}
\usepackage{subfigure}
\usepackage{fancyhdr}	

\usepackage{emptypage}
\usepackage[english]{varioref}
\usepackage{color}
\usepackage{eurosym}
\usepackage{lettrine}

\usepackage{amsmath}
 
\DeclareMathOperator{\diag}{diag}
\usepackage{mathrsfs}
\usepackage{amsfonts}
\usepackage{amssymb}
\usepackage{bm}
\usepackage{etoolbox} 
\newtheorem{theorem}{Theorem}
\newtheorem{proposition}{Proposition}

\newtheorem{objective}{Objective}
\newtheorem{remark}{Remark}
\newtheorem{assumption}{Assumption}
\usepackage{algorithm}
\usepackage{dsfont}
\usepackage{algpseudocode}
\usepackage{upgreek}
\newcommand{\numberset}{\mathbb}

\newcommand{\R}{\numberset{R}}

\usepackage{mathtools}

\usepackage{amscd}
\usepackage{booktabs}
\usepackage{graphics}
\usepackage{epsfig}
\usepackage{contour}
\usepackage{bm}
\usepackage{times}
\usepackage{rotating}
\usepackage{tabularx}
\usepackage{longtable}
\usepackage{pdflscape}
\usepackage{cite}
\usepackage{listings}
\usepackage{xcolor}
\usepackage{siunitx}
\renewcommand{\vec}{\bm}
\renewcommand{\thefootnote}{\fnsymbol{footnote}}

\usepackage{cases}
\usepackage{cite}
\usepackage{siunitx}	
\usepackage{tikz}
\usetikzlibrary{arrows,automata}
\usetikzlibrary{arrows,shapes,backgrounds,calc,positioning,patterns}
\usepackage{balance}
\usetikzlibrary{calc,arrows,shapes,backgrounds,calc,positioning,patterns,decorations.pathmorphing,decorations.markings,mindmap,trees}
\tikzstyle{block} = [draw, rectangle, minimum height=2em, minimum
width=4em] \tikzstyle{sum} = [draw, fill=blue!20, circle, node
distance=1cm] \tikzstyle{input} = [coordinate] \tikzstyle{output} =
[coordinate] \tikzstyle{pinstyle} = [pin edge={to-,thin,black}]
\usepackage{blox}
\usepackage{cases}
\usepackage{framed}
\colorlet{shadecolor}{black!15}
\usepackage{bigints}
\usepackage[american,cute inductors,smartlabels]{circuitikz}
\usetikzlibrary{arrows,automata}
\usepackage[american,cuteinductors,smartlabels]{circuitikz}

\usetikzlibrary{calc}
\ctikzset{bipoles/thickness=1} \ctikzset{bipoles/length=0.8cm}
\ctikzset{bipoles/diode/height=.375}
\ctikzset{bipoles/diode/width=.3}
\ctikzset{tripoles/thyristor/height=.8}
\ctikzset{tripoles/thyristor/width=1}
\ctikzset{bipoles/vsourceam/height/.initial=.7}
\ctikzset{bipoles/vsourceam/width/.initial=.7} \tikzstyle{every
node}=[font=\small] \tikzstyle{every path}=[line width=0.8pt,line
cap=round,line join=round]
\newcommand{\ncom}{\newcommand}
\ncom{\beqn}{\begin{eqnarray*}} \ncom{\eeqn}{\end{eqnarray*}}
\ncom{\beq}{\begin{eqnarray}} \ncom{\eeq}{\end{eqnarray}}

\begin{document}
\begin{frontmatter}

	\title{Voltage control of DC networks: robustness for unknown ZIP-loads}  
	
	\thanks[footnoteinfo]{M. Cucuzzella and K. C. Kosaraju have given equivalent contribution.}
	\thanks[footnoteinfo]{This work is  supported by the EU Project `MatchIT' (project number: 82203) and the Netherlands Organisation for Scientific Research through Research Programme ENBARK+ under Project 408.urs+.16.005.}
	
	\author[Paestum]{Michele Cucuzzella}$^\star$\ead{m.cucuzzella@rug.nl},    % Add the 
	\author[Paestum]{Krishna Chaitanya Kosaraju}$^\star$\ead{k.c.kosaraju@rug.nl},               % e-mail address 
	\author[Paestum]{Jacquelien M. A. Scherpen}\ead{j.m.a.scherpen@rug.nl}  % (ead) as shown
	
	\address[Paestum]{Jan C. Wilems Center for Systems and Control, ENTEG, Faculty of Science and Engineering, University of Groningen, Nijenborgh 4, 9747 AG Groningen, the Netherlands.}

	                             % keyword list or with the 

	\begin{abstract}
	In this paper we propose a new passivity-based control technique for DC power networks comprising the so-called ZIP-loads, i.e., nonlinear loads with the parallel combination of unknown constant impedance (Z), current (I) and power (P) components. 
	More precisely, we propose a novel passifying input and a storage function based on the so-called mixed potential function introduced by Brayton and Moser, leading to a novel passivity property with  output port-variable equal to the first time derivative of the voltage. 
	Differently from the existing results in the literature, where restrictive (sufficient) conditions on Z, P and the voltage reference are assumed to be satisfied, we establish a passivity property for every positive voltage reference and every type of load.  
	Consequently, we develop a new decentralized passivity-based control scheme that is robust with respect to the uncertainty affecting the ZIP-loads. 
	\end{abstract}
	
\end{frontmatter}

%========================================INTRODUCTION

\section{Introduction}

Loads in a power network can be broadly classified into two groups: \emph{nonactive} and \emph{active} loads. Common examples for nonactive loads are constant impedance (Z) and constant current (I) loads. More precisely, the voltage across a Z-load and the corresponding absorbed current satisfy a  positive linear relationship, while an I-load varies its internal impedance in order to absorb a constant positive current regardless of the applied voltage. 
However, due to the advancement in power electronics in the past decade, a considerable percentage of the loads consists of active loads (e.g., motor drives, power converters and electronic devices), which often behave as constant power (P) loads (see for instance~\cite{en10101656,SINGH2017407} and the references therein). Specifically, independently of the applied voltage, a P-load varies its internal impedance in order to absorb a constant positive power (i.e., it has an intrinsic negative incremental impedance). Unfortunately, the presence of negative incremental impedance may cause voltage oscillations that impact the power quality or even lead to the network instability~\cite{en10101656,1461638}. 
Motivated by this and the growing research interest on Direct-Current (DC) microgrids (see for instance~\cite{7268934,7890986,DePersis2016,J_Trip2018} and the references therein), the main focus of this paper is the design and analysis of a robust control technique for DC networks comprising \emph{unknown} ZIP-loads. 
Microgrids are indeed (low-voltage) electricity networks, wherein loads, sources and storage units require careful coordination \cite{C_Lasseter_14}. Generally, microgrids can be Alternating-Current (AC) or DC networks. 
However, the advancement in power electronics technology together with the increasing number of DC sources and loads (e.g. photovoltaic panels, batteries, electronic appliances) are moving the interest towards DC microgrids, which are in some situations more efficient and reliable than AC microgrids \cite{Justo2013387}.

\subsection{Literature review}
The main control objective in (islanded) DC networks is voltage regulation, guaranteeing the network stability and a proper functioning of the connected loads~\cite{7268934,7890986}. In the last decades, several controllers based on different techniques have been proposed in the literature, e.g., droop~\cite{ZHAO201518}, plug-and-play~\cite{7934339,7983406,StrehlePfeiferKrebs2019_1000090285}, sliding mode~\cite{cucuzzella2017robust,doi:10.1080/00207179.2017.1306112} and passivity-based~\cite{Sira,1323174} controllers. However, these works do not include P-loads and, some of them, analyze the stability of the single converter or assume the network to be purely resistive.

In order to address the voltage destabilizing effect of the negative incremental impedance introduced by P-loads, several approaches have been proposed in the literature. 
A simple solution is to (sufficiently) increase the damping of the system by connecting a positive impedance in parallel to the P-load. In the literature, this technique is called {\em passive-damping} (see for instance~\cite{5764841,7402398} and the references therein). 
However, besides increasing the size, weight and cost of the system, this method requires extra power for supplying the added impedance.
As an alternative to the passive-damping technique, it is generally possible to design a feedback control that produces the effect of a virtual resistor and damps the voltage oscillations. This methodology is called {\em active-damping} (see for instance~\cite{Sira,1323174,4776490,6506754,WU2014,7492256} and the references therein). 

More recently, energy-based approaches have been proposed in~\cite{DePersis2016,DBLP:journals/corr/abs-1802-02483,253266,machado2018active} to analyze DC networks including ZIP-loads. More specifically, in~\cite{DePersis2016}, the authors propose a consensus algorithm that is analyzed via Lyapunov functions inspired by the physics of the system, guaranteeing power consensus and preserving the geometric mean of the source voltages. In~\cite{DBLP:journals/corr/abs-1802-02483}, the authors provide a suitable port-Hamiltonian framework to model electrical circuits including ZIP-loads and investigate their shifted passivity properties. In~\cite{253266}, the authors show that the controllers proposed in~\cite{7934339} passivate the generation and load units of a DC microgrid. However, (sufficient) conditions on Z, P and the voltage reference are assumed to be satisfied (see~\cite[Theorem 1]{DePersis2016}, \cite[Section 5]{DBLP:journals/corr/abs-1802-02483} and \cite[Theorem 2]{253266}, respectively). In~\cite{machado2018active}, the authors propose an adaptive passivity-based control scheme that include an estimator of the power consumed by the P-load. 

\subsection{Main contributions}
In this paper, we use passivity theory~\cite{HILL1980327,915398,van2000l2} to design and analyze a decentralized robust control scheme for DC networks comprising \emph{unknown} ZIP-loads.
More precisely, inspired by the theory developed in~\cite{brayton1964theorya,brayton1964theoryb,1230225, 1235380}, we propose a novel passifying input and a storage function that lead, under a very mild assumption, to the establishment of a passivity property where the output port-variable is equal to the first time derivative of the voltage. Specifically, the passifying input includes (i) a feedforward action to compensate the voltage drop on the filter resistance and (ii) a damping injection by modification of the dissipative structure of the system. Specifically, the latter action is equivalent to add  a (nonlinear) \emph{virtual} resistor connected in parallel to the \emph{real} P-load to compensate its negative incremental impedance.
Then, we shape the closed-loop storage function such that it has a minimum at the desired operating point, where the network voltage is equal to the corresponding reference value.

%Moreover, differently from the existing results in the literature, where restrictive (sufficient) conditions on the load parameters and voltage reference are assumed to be satisfied, the considered DC network in closed loop with the proposed passifying input is passive for all the trajectories evolving in the subspace of the state-space where the voltage is positive.

We now list the main contributions of this work:
\begin{itemize}
\item[(1)]{\em Robustness}: The proposed control scheme is decentralized, scalable and  robust with respective to the uncertainty affecting the ZIP-loads and the power lines impedances. Moreover, the proposed control scheme does not require the implementation of observers to estimate the actual power absorbed by the P-loads.\\

	\item [(2)]{\em Passivity}: We establish a passivity property with the output port-variable equal to the first time derivative of the voltage. As a consequence, we simply shape the closed-loop storage function (introducing a function of the voltage) such that it has a minimum at the desired operating point. More precisely, we adopt the {\em output-shaping} technique introduced in~\cite{2018arXiv181102838C}, where the integrated passive output is used to shape the closed-loop storage function.\\
	
	\item[(3)]{\em Less restrictive conditions}: Differently from most of the existing results in the literature, where restrictive (sufficient) conditions on the load parameters and voltage reference are assumed to be satisfied, the considered DC network in closed loop with the proposed passifying input is passive for all the trajectories evolving in the subspace of the state-space where the voltage is positive (for every positive voltage reference and every type of load, even loads consisting of only the P component).
\end{itemize}

%========================================

\subsection{Organization of the paper} 
The paper is organized as follows. In Section \ref{sec:Gen_BM}, we review the Brayton-Moser (BM) formulation of resistive-inductive-capacitive (RLC) circuits and present a preliminary result, i.e., a novel family of BM descriptions and its passivity property, which is fundamental for establishing later the passivity property for the considered DC network. In Subsection \ref{subsec:model}, we present the model of a DC network including ZIP-loads and show that it can be expressed as a BM system. Furthermore, in Subsection \ref{subsec:prob_form}, we formulate the voltage regulation problem. In Subsection~\ref{subsec:exist_passivity_properties}, we present and discuss some of the passivity properties  proposed in the literature for DC networks, putting in evidence the corresponding limitations. In Subsection~\ref{subsec:passivity}, we present our first main result, i.e., a novel passivity property for DC networks including unknown ZIP-loads. 
Based on the established passivity property, in Section \ref{sec:pbc}, we present the second main result, i.e., the design and analysis of a new passivity-based control scheme that robustly stabilizes the DC network voltage to the corresponding desired value. 
Finally, we test our controller numerically in Section \ref{sec:sim}, and gather some conclusions and future research directions in Section \ref{sec:con}. 
           \subsection{Notation}
           The set of real numbers and strictly positive real numbers are denoted by $\mathbb{R}$ and $\mathbb{R}_{>0}$, respectively. For a vector $x \in \mathbb{R}^n$ and a symmetric and positive semidefinite matrix~$M \in \mathbb{R}^{n \times n}$, let~$\| x\|_M:= (x^\top M x)^{1/2}$. If~$M$ is the identity matrix, then the Euclidean norm is denoted by~$\|x\|$. 
           Given a function $f:\mathbb{R}^n\rightarrow \mathbb{R}$,  $\nabla_xf(x) \in \mathbb{R}^n$ denotes the partial derivatives of $f(x)$ with respect to $x$, i.e., $\nabla_xf(x):=(\partial f/\partial x_1, \cdots, \partial f/\partial x_n)^{\top}$.
           For symmetric matrices $P,Q \in \mathbb{R}^{n\times n}$, $P \le Q$ implies that $Q-P$ is positive semidefinite. $\mathcal{I}_n$ represents the identity matrix of order $n$, while \lq$\mathrm{1}$' denotes the ones vector of appropriate dimension. The steady state solution to the system $\dot{x}=\zeta(x)$, is denoted by $\overline{x}$, i.e., ${\bf 0}=\zeta(\overline{x})$. A constant signal is denoted by $~x^*$. Let $v\in \mathbb{R}^n$, then $[v]:=\diag\{v_1,\cdots, v_n\}$, and $\ln v :=[\ln v_1, \cdots, \ln v_n]^\top$. The argument of a function is omitted if it is clear from the context.
%
%
%
%
%======================================== BM

    \section{Brayton-Moser framework}\label{sec:Gen_BM} 
    In the early 1960s, Brayton and Moser developed an unified framework where the dynamics of a class of RLC electrical circuits are described by a gradient structure~\cite{brayton1964theorya,brayton1964theoryb}. 
    In the literature, this property has been widely used for controlling and analyzing RLC circuits (see for instance~\cite{1230225, 1235380,1323174,5184955,7798761,2018arXiv181102838C,8619192} and the references therein).

   In the next two subsections, for the readers' convenience, we recall some basic notions of the BM theory and its generalization through the generation of a family of BM descriptions, which will be used in Subsection \ref{subsec:passivity} to establish a novel passivity property for DC networks including ZIP-loads.

%======================================== 
   
    \subsection{Preliminaries}
       Consider the class of {topologically complete} RLC circuits~\cite{1393170} with $\sigma$ inductors, $\rho$ capacitors {and $k\leq \sigma$ (current-controlled) voltage sources  $u: \mathbb{R}_{\geq0} \rightarrow \mathbb{R}^k$ connected in series with the inductors}. In~\cite{brayton1964theorya,brayton1964theoryb}, Brayton and Moser show that the dynamics\footnote{For notational simplicity, the dependency of the variables on time is mostly omitted throughout this paper.} of this class of systems can be represented as follows:
\begin{align}
\label{BM_non_switched}
\begin{split}
-L\dot{I}&=\nabla_I\mathcal{P}(I,V) + Bu\\
C\dot{V}&=\nabla_V\mathcal{P}(I,V),
\end{split}
\end{align}
where $L\in \mathbb{R}^{\sigma \times \sigma}$ and $C\in
\mathbb{R}^{\rho \times \rho}$ are positive definite symmetric
matrices with the inductances and capacitances as  entries,
respectively. The signals $I: \mathbb{R}_{\geq0} \rightarrow\mathbb{R}^{\sigma}$ and $V: \mathbb{R}_{\geq 0} \rightarrow \mathbb{R}^{\rho}$ denote the currents through the $\sigma$
inductors and the voltages across the $\rho$ capacitors,
respectively. The matrix $B\in \mathbb{R}^{\sigma\times k}$ is the
input matrix and $\mathcal P:\mathbb{R}^{\sigma }\times\mathbb{R}^{ \rho}\rightarrow
\mathbb{R}$ represents the so-called \emph{mixed potential}
function, which can be expressed as follows:
\begin{equation}
\label{Mixpot}
\mathcal{P}(I,V) =I^\top\Gamma V+F(I)-\mathcal{G}(V),
\end{equation}
where the matrix {$\Gamma\in \mathbb{R}^{\sigma\times \rho}$} captures the instantaneous power transfer between the storage elements (i.e., inductors and capacitors). The resistive content $F:\mathbb{R}^\sigma \rightarrow \mathbb{R}$ and the resistive co-content $\mathcal G:\mathbb{R}^\rho \rightarrow \mathbb{R}$ capture the power dissipated for instance in the resistors  connected in series to the inductors and in parallel to the capacitors, respectively.  
Compactly, the BM equations  \eqref{BM_non_switched} can be expressed as follows:
\begin{equation}\label{gradient_structure}
Q\dot{x}= \nabla_x\mathcal{P}(x)+\tilde{B}u,
\end{equation}
where $x=(I^\top,\;V^\top)^\top$, $Q=\diag\{-L,C\}$ and $\tilde{B}=(B^\top,~\mathcal{O}_{k\times \rho })^\top$.

%========================================

\subsection{Generalized gradient structure}\label{subsec:gradient_structure}

The mixed potential function \eqref{Mixpot} satisfies 
\begin{align}\label{gen_BM_passivity}
\begin{split}
	\dot{\mathcal P}(x)&=\nabla_x^\top \mathcal{P}(x)\dot{x}\\ 
	&=\left(Q\dot{x}-\tilde{B}u\right)^\top \dot{x}\\
	&=\dfrac{1}{2}\|\dot{x}\|_{\left(Q+Q^\top \right)}^2-\dot{x}^\top\tilde{B} u,
	\end{split}
\end{align}
along the solutions to \eqref{gradient_structure}.
Therefore, \eqref{gen_BM_passivity} implies that system~\eqref{gradient_structure} is passive with supply rate  $-\dot{x}^\top\tilde{B}^\top u$, if the mixed potential function $\mathcal P$ is positive semi-definite and the matrix $Q$ is negative semi-definite.
 Unfortunately, the class of systems that satisfies this property is small and restricted for instance to  RL or RC circuits~\cite{5184955}. For the considered class of systems \eqref{gradient_structure}, it is indeed straightforward to verify that the symmetric part of $Q$ is indefinite.
 However, in~\cite{brayton1964theorya}, Brayton and Moser observed that it is possible to generate a new pair $\{Q_A, \mathcal{P}_A\}$ that preserves the gradient structure \eqref{gradient_structure}. 
 
In this paper, differently from~\cite{1230225, 1235380}, we propose a \emph{novel} family of BM descriptions and provide its complete characterization in the following proposition.

\begin{proposition}\emph{(A novel family of BM descriptions).}
	\label{lemma:gradient_generalized}
	For $\lambda \in \mathbb{R}$, full rank matrix $D\in \R^{k\times k}$, constant symmetric matrix $M\in \mathbb{R}^{(\sigma+\rho)^2}$ and $Q_0:\mathbb{R}^{\sigma+\rho} \rightarrow \mathbb{R}^{k \times (\sigma+\rho)}$, system~\eqref{gradient_structure} can be (re)written as follows: 
	\begin{equation}\label{gradient_generalized}
	Q_A(x)\dot{x}=\nabla_x\mathcal{P}_A(x)+\tilde{B}_A(x)\upsilon,
	\end{equation}
	where
	\begin{subequations}\label{generalized_BM}
		\begin{align}
		Q_A(x)&:=(\lambda \mathcal{I}_{\sigma+\rho}+\nabla_x^2\mathcal{P}(x)M)(Q-\tilde{B}Q_0(x))\label{generalized_BMa}\\
		\mathcal{P}_A(x)&:=\lambda \mathcal P(x)+\dfrac{1}{2}\nabla_x^\top\mathcal{P}(x) M\nabla_x\mathcal{P}(x)\label{generalized_BMb}\\
		\tilde{B}_A(x)&:=(\lambda \mathcal{I}_{\sigma+\rho}+\nabla_x^2\mathcal{P}(x)M)\tilde{B}D\label{generalized_BMc}\\
		\upsilon&:=D^{-1}\left(u-Q_0(x)\dot{x}\right),\label{generalized_BMd}
		\end{align}
	\end{subequations}
	with $\lambda$ and $M$ such that $(\lambda \mathcal{I}_{\sigma+\rho}+\nabla_x^2\mathcal{P}(x)M)$ has full rank.
\end{proposition}
\begin{pf}
	A straightforward computation shows that the solution to \eqref{gradient_structure} precisely coincides with the solution to \eqref{gradient_generalized}, i.e.,
\begin{align*}
\begin{split}
Q^{-1}\left(\nabla_x\mathcal{P}(x)+\tilde{B}u\right) =& \left(Q_A(x)+\tilde{B}_A(x)D^{-1}Q_0(x)\right)^{-1}\\
&\cdot \left(\nabla_x\mathcal{P}_A(x)+\tilde{B}_A(x)D^{-1}u\right).
\end{split}
\end{align*}
	Moreover, $\mathcal{P}_A$ in \eqref{generalized_BMb} satisfies
		\begin{align*}
		\begin{split}
		\nabla_x\mathcal{P}_A(x)&=(\lambda \mathcal{I}_{\sigma+\rho}+\nabla_x^2\mathcal{P}(x)M)\nabla_x\mathcal{P}(x)\\
%		&= (\lambda \mathcal{I}_{\sigma+\rho}+\nabla_x^2\mathcal{P}M)(Q\dot{x}-\tilde{B}u)\\
%		&= (\lambda \mathcal{I}_{\sigma+\rho}+\nabla_x^2\mathcal{P}M)((Q-\tilde{B}Q_0)\dot{x}-\tilde{B}D\upsilon)\\
		&= Q_A(x)\dot{x}-\tilde{B}_A(x)\upsilon.
		\end{split}
		\end{align*}
		$\hfill\blacksquare$
\end{pf} 

\begin{remark} \emph{(Novelty of \eqref{gradient_generalized}).}
	Note that the particular structure of $\upsilon$ that we propose in~\eqref{generalized_BMd} generates a family of BM descriptions \eqref{gradient_generalized} different from the ones presented for instance in~\cite{1230225, 1235380}. The proposed structure plays indeed a major role in establishing the (novel) passivity property for DC networks including unknown ZIP-loads (see Theorem \ref{thm:proposed_passivity} in Subsection \ref{subsec:passivity}). More precisely, the term $Q_0 \dot{x}$ in~\eqref{generalized_BMd} is essential to counteract the effect of the P component.
\end{remark}

As a direct consequence of Proposition \ref{lemma:gradient_generalized}, and in analogy with~\cite{1230225, 1235380}, the following preliminary result is presented.
\begin{proposition} \emph{(Passivity property of~\eqref{gradient_generalized}).}
Assume that the pair $\{Q_A, \mathcal{P}_A\}$ in \eqref{generalized_BM} satisfies $\mathcal{P}_A\geq0 $ and $Q_A+Q_A^\top\leq 0$, for all $x\in\R^{\sigma + \rho}$. Then, system~\eqref{gradient_generalized} is passive with respect to the storage function  $\mathcal{P}_A$ and supply rate $-\dot{x}^\top\tilde{B}_A\upsilon$.
\label{prop:2}
\end{proposition}
\begin{pf}
Similarly to \eqref{gen_BM_passivity}, the function $\mathcal{P}_A$ in \eqref{generalized_BMb} satisfies 
\begin{align*}
\begin{split}
\dot{\mathcal P}_A(x)%&=\nabla_x^\top \mathcal{P}_A\dot{x}\\
%&=\left(Q_A\dot{x}-\tilde{B}_A\upsilon\right)^\top \dot{x}\\
%&=\dfrac{1}{2}\|\dot{x}\|_{\left(Q_A(x)+Q^\top_A(x) \right)}^2-\dot{x}^\top\tilde{B}_A(x) \upsilon(x,u)\\
&\leq -\dot{x}^\top\tilde{B}_A(x)\upsilon,
\end{split}
\end{align*}
along the solutions to \eqref{gradient_generalized}.
$\hfill\blacksquare$
\end{pf}
%
%
%
%
%======================================== DC 

\section{DC power network} 
    \label{sec:model}
    \begin{figure}[t]
    	\begin{center}
    		\begin{circuitikz}[scale=.95,transform shape]
    			\ctikzset{current/distance=1}
    			\draw
    			% transformators i and j
    			node[] (Ti) at (0,0) {}
    			node[] (Tj) at ($(5.4,0)$) {}
    			% Buck i
    			node[] (Aibattery) at ([xshift=-4.5cm,yshift=0.9cm]Ti) {}
    			node[] (Bibattery) at ([xshift=-4.5cm,yshift=-0.9cm]Ti) {}
    			node[] (Ai) at ($(Aibattery)+(0,0.2)$) {}
    			node[] (Bi) at ($(Bibattery)+(0,-0.2)$) {}
    			(Ai) to [R, l={$R_{si}$}] ($(Ai)+(1.7,0)$) {}
    			($(Ai)+(1.7,0)$) to [short,i={$I_{si}$}]($(Ai)+(1.701,0)$){}
    			($(Ai)+(1.701,0)$) to [L, l={$L_{si}$}] ($(Ai)+(3,0)$){}
    			to [short, l={}]($(Ti)+(0,1.1)$){}
    			(Bi) to [short] ($(Ti)+(0,-1.1)$);
    			\draw
    			($(Ai)$) to []($(Aibattery)+(0,0)$)to [V_=$u_i$]($(Bi)$)
    			% PCC-i
    			($(Ti)+(-1.3,1.1)$) node[anchor=south]{{$V_{i}$}}
    			($(Ti)+(-1.3,1.1)$) node[ocirc](PCCi){}
    			($(Ti)+(-.3,1.1)$) to [short,i>={$I_{li}(V_i)$}]($(Ti)+(-.3,0.5)$)to [I]($(Ti)+(-.3,-1.1)$)
    			($(Ti)+(-1.3,1.1)$) to [C, l_={$C_{si}$}] ($(Ti)+(-1.3,-1.1)$)
    			% line
    			($(Ti)+(2.,1.1)$) to [short,i={$I_{tk}$}] ($(Ti)+(2.2,1.1)$)
    			($(Ti)+(0,1.1)$)--($(Ti)+(.6,1.1)$) to [R, l={$R_{tk}$}] 
    			($(Ti)+(2.5,1.1)$) {} to [L, l={{$L_{tk}$}}, color=black]($(Tj)+(-2.2,1.1)$){}
    			($(Tj)+(-2.2,1.1)$) to [short]  ($(Ti)+(3.4,1.1)$)
    			($(Ti)+(0,-1.1)$) to [short] ($(Ti)+(3.4,-1.1)$);
    			\draw
    			node [rectangle,draw,minimum width=6.1cm,minimum height=3.4cm,dashed,color=gray,label=\textbf{DGU $i$},densely dashed, rounded corners] (DGUi) at ($0.5*(Aibattery)+0.5*(Bibattery)+(2.25,0.2)$) {}
    			node [rectangle,draw,minimum width=2.2cm,minimum height=3.4cm,dashed,color=gray,label=\textbf{Line $k$},densely dashed, rounded corners] (DGUi) at ($0.5*(Aibattery)+0.5*(Bibattery)+(6.65,0.2)$) {};
    		\end{circuitikz}
    		\caption{Electrical scheme of DGU $i$ and transmission line $k$.}
    		\label{fig:networks}
    	\end{center}
    	\vspace{.2cm}
    \end{figure}
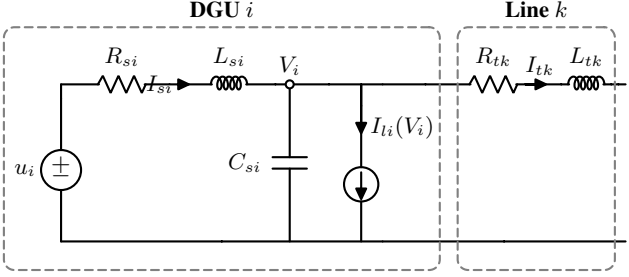
    \begin{table}
	\begin{center}
		\caption{Description of the used symbols}\label{tab:symbols2}
		\vspace{.2cm}
		\begin{tabular}{cl}
			\toprule
			& {\bf State variables} \\
			\midrule
			$I_{si}$						& Generated current\\
			$V_i$						& Load voltage\\
			$I_{tk}$ 						& Line current \\
			\midrule
			& {\bf Input}\\
			\midrule
			$u_i$						& Control input \\
			\midrule
			& {\bf Parameters}\\
			\midrule
			$L_{si}$						& Filter inductance\\
			$C_{si}$						& Filter capacitor\\
			$R_{si}$						& Filter resistance\\
			$R_{tk}$						& Line resistance\\
			${L_{tk}}$						& {Line inductance}\\
			\midrule
			& {\bf Load}\\
			\midrule
			$Z_{li}^*$						& Constant impedance\\
			$I_{li}^*$						& Constant current\\
			$P_{li}^*$						&constant power\\
			\bottomrule
		\end{tabular}
	\end{center}
		\vspace{.2cm}
\end{table}
    In this section, we present the BM formulation of a DC network composed of $n$ Distributed Generation Units (DGUs) connected to each other through $m$ RL power lines. Then, after formulating the control objective, we introduce the only assumption of the paper, motivating its plausibility (see Remark~\ref{rm:practical_considerations_Pi}) and discussing its  technical implications (see Remark~\ref{rm:technical_considerations_Pi}).

%======================================== 

    \subsection{Model}\label{subsec:model}
    A schematic electrical diagram of the considered DC network including a DGU and a transmission line  is illustrated in Fig.~\ref{fig:networks} (see also Table~\ref{tab:symbols2} for the description of the used symbols).
    Each DGU represents for instance a DC-DC buck converter (including an output RLC low-pass filter) supplying an unknown load.
    By using the Kirchhoff's current and voltage laws, the  equations describing the dynamic behavior of the DGU $i$ are given by
    \begin{align}
    \begin{split}
    \label{eq:plant_i1}
    -L_{si}\dot{I}_{si} &=   R_{si}I_{si}+V_i - u_i \\
    C_{si}\dot{V}_i &=  I_{si} - I_{li}(V_i) -  \displaystyle{ \sum_{k \in \mathcal{E}_i}^{}I_{tk}},
    \end{split}
    \end{align}
    where $I_{si}: \R_{\geq0} \rightarrow \R, V_i:\R_{\geq0} \rightarrow \R, I_{li}: \R \rightarrow \R_{\geq0}, I_{tk}:\R_{\geq0} \rightarrow \R, u_i : \R_{\geq0} \rightarrow \R$ and $L_{si}, C_{si} \in \R_{>0}$. Moreover, $\mathcal{E}_i$ is the set of  lines connected to the DGU~$i$. The control input $u_i$ represents for instance the output voltage of a buck converter.	The current shared among DGU $i$ and DGU $j$ through the  line $k$ is denoted by $I_{tk}$, and its dynamic is given by
    \begin{equation}
    {
    	\label{eq:plant_i2}
    	-L_{tk}{\dot I_{tk}} = R_{tk} I_{tk}+(V_j - V_i), }
    \end{equation}
    where $L_{tk}, R_{tk}\in \mathbb{R}_{>0}$. Moreover, the term $I_{li}(V_i)$ in \eqref{eq:plant_i1} represents the \emph{unknown} current demanded by load $i \in \mathcal{V}$ and (generally) depends on the node voltage $V_i$. 
    
\begin{figure}[]
\begin{center}
\hspace{-3cm}
\begin{circuitikz}[scale=0.95,transform shape]
\ctikzset{current/distance=1}
\draw
node[] (Ti) at (0,0) {}
node[] (Tj) at ($(5.4,0)$) {}
($(Tj)+(-1.8,1.1)$) to [short] ($(Tj)+(-1.2,1.1)$)--($(Tj)+(-0.8,1.1)$)
($(Tj)+(-0.8,1.1)$) to [short] ($(Tj)+(2.6,1.1)$)
($(Tj)+(-1.8,-1.1)$) to [short] ($(Tj)+(2.6,-1.1)$)
% PCC-j
($(Tj)+(-1.8,1.1)$) node[ocirc](PCCj){}
($(Tj)+(-1.8,-1.1)$) node[ocirc](PCCjs){};
\begin{scope}[shorten >= 10pt,shorten <= 10pt,]
\draw[<-] ($(Tj)+(-1.8,1.1)$) -- node[right] {$V_{i}$} ($(Tj)+(-1.8,-1.1)$);
\end{scope};
\draw
($(Tj)+(-1.75,1.1)$) to [short,i={$I_{li}(V_i)$}]($(Tj)+(-.8,1.1)$){}
($(Tj)+(1.4,1.1)$)--($(Tj)+(1.4,0.8)$) to [short,i>={$I_{li}^\ast$}]($(Tj)+(1.4,0.5)$)to [I]($(Tj)+(1.4,-1.1)$)
($(Tj)+(2.6,1.1)$)--($(Tj)+(2.6,0.8)$) to [short,i>={$\dfrac{P_{li}^\ast}{V_{i}}$}]($(Tj)+(2.6,0.5)$)to [I]($(Tj)+(2.6,-1.1)$)
($(Tj)+(0.2,1.1)$) to [R, l={$Z^{*}_{li}$}] ($(Tj)+(0.2,-1.1)$);
\end{circuitikz}
\caption{{ZIP-load $i$.}}
\label{fig:ZIPload}
\end{center}
	\vspace{.2cm}
\end{figure}
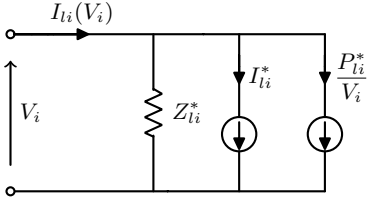

    In this work, we consider a general nonlinear load model including the parallel combination of the following load components~(see Figure \ref{fig:ZIPload} for the corresponding circuit representation):
    \begin{enumerate}
    	\item{constant impedance: $I_{li} = Z^{*-1}_{li}V_i$, with $Z^{*-1}_{li}\in\R_{\geq 0}$,}
    	\item{constant current: $I_{li} = I^*_{li}$, with $I^*_{li}\in\R_{\geq0}$, and}
    	\item{constant power: $I_{li} = V_i^{-1}P^*_{li}$, with $P^*_{li}\in\R_{\geq0}$.}
    \end{enumerate}
     Accordingly, in the presence of a ZIP-load, $I_{li}(V_i)$ in \eqref{eq:plant_i1} is given by
    \begin{align}\label{eq:plant_i3}
    	I_{li}(V_i):=Z^{*-1}_{li}V_i+I^*_{li}+V_i^{-1}P^*_{li}.
    \end{align}
    The symbols used in  \eqref{eq:plant_i1}--\eqref{eq:plant_i3} are described in Table~\ref{tab:symbols2}.
 
    The overall DC network is represented by a connected and undirected graph $\mathcal{G} = (\mathcal{V},\mathcal{E})$, where the nodes, $\mathcal{V} = \{1,...,n\}$, represent the DGUs and the edges, $\mathcal{E}  = \{1,...,m\}$, represent the  lines interconnecting the DGUs.
    The network topology is described by its corresponding incidence matrix $\mathcal{B} \in \R^{n \times m}$. The ends of edge $k$ are arbitrarily labeled with a $+$ and a $-$, and the entries of  $\mathcal{B}$ are given by
    \begin{equation*}
    \mathcal{B}_{ik}=
    \begin{cases}
    +1 \quad &\text{if $i$ is the positive end of $k$}\\
    -1 \quad &\text{if $i$ is the negative end of $k$}\\
    0 \quad &\text{otherwise}.
    \end{cases}
    \end{equation*}
    Consequently, the overall dynamical system describing the network behavior can be written compactly for all the DGUs $i \in \mathcal{V}$ as follows:
    {
    	\begin{align}
    	\label{eq:plant_mg}
    	\begin{split}
    	-L_s\dot{I}_{s} & =R_{s}I_{s}+V - u\\
    	-	L_t\dot{I}_t & = R_t I_t +\mathcal{B}^\top V\\
    	C_{s}\dot{V} & = I_{s} + \mathcal{B}I_t - I_l(V),		
    	\end{split}
    	\end{align}
    }where $I_s:\R_{\geq 0}\rightarrow \R^{n}$, {$I_t : \R_{\geq 0}\rightarrow \R^m$}, $V:\R_{\geq 0}\rightarrow \R^{n}$, $u:\R_{\geq 0}\rightarrow \R^{n}$, and $I_l:\R^n \rightarrow \R^n_{\geq0}$, with $P_l^*, ~I_l^*\in \R^n_{\geq0}$.
    Moreover, $C_s, L_s, R_s, Z_l^*\in \R^{n \times n}$ and $R_t, {L}_t \in \R^{m \times m}$ are positive definite diagonal matrices, e.g., $C_s = \diag \{C_{s1}, \cdots, C_{sn}\}$.

The DC network \eqref{eq:plant_mg} can be written in the BM structure~\eqref{gradient_structure} using $\sigma=n+m$, $\rho=n$, $k=n$,  $I=[I_s^\top,I_t^\top]^\top$, $x=[I_s^\top, I_t^\top, V^\top]^\top$, $L=\diag\{L_s,L_t\}$, $C=C_s$, and
\begin{subequations}\label{BM_mg}
\begin{align}
Q&=\diag\{-L_s,-L_t,C_s\}\label{BM_mga}\\
\Gamma&=[\mathcal{I}_n \;\;\mathcal{B}]^\top\label{BM_mgb}\\
F(I)&=\dfrac{1}{2}\|I_t\|_{R_t}^2+\dfrac{1}{2}\|I_s\|_{R_s}^2\label{BM_mgc}\\
\mathcal{G}(V)&=\dfrac{1}{2}\|V\|_{Z_l^{*-1}}^2+P_l^{*\top}\ln{V}+I_l^{*\top} V\label{BM_mgd}\\
\tilde{B}&=[-\mathcal{I}_n\;\; \mathcal{O}_{n \times m}\;\;\mathcal{O}_n]^\top.\label{BM_mge}
\end{align}
\end{subequations}

We recall now a well-known Bryaton and Moser's result concerning the stability of system~\eqref{eq:plant_mg}. 
  \begin{theorem}\emph{(~\cite[Theorem 3]{brayton1964theorya}~).}
  	System \eqref{eq:plant_mg} with $u=u^*\in \R^n_{>0}$ is (non-oscillatory) asymptotically stable  if $R_s, R_t$ are positive definite, $P_l^{*\top}\ln V+ \|\Gamma V\|\rightarrow \infty$ as $\|V\|\rightarrow \infty$ and 
  	\begin{align}\label{BM_M_original_cond}
  	\|L^{1/2}\diag\{R_s^{-1},R_t^{-1}\}\Gamma C^{-1/2}_s\|\leq 1-\delta, \quad \delta>0.
  	\end{align}
  \end{theorem}
  
We show in Sections~\ref{sec:passivity} and~\ref{sec:pbc} that the proposed control strategy asymptotically stabilizes system \eqref{eq:plant_mg} independently from condition \eqref{BM_M_original_cond}.

%========================================

\subsection{Problem formulation}
\label{subsec:prob_form}
In DC networks, the main control objective is to regulate the voltage across the load towards a desired reference value. Before introducing the control objective, we first show that for every constant input $u= u^\ast \in \R^n$, the steady state solution to system \eqref{eq:plant_mg} is the following:
	\begin{subequations}
	\begin{align}
	\overline{V} & =  -R_s\bar{I}_s+{u^\ast} \label{eq:ssV}\\
	\overline I_t & = -R^{-1}_t\mathcal{B}^\top\overline V  \label{eq:ssIt}\\
	\overline{I}_{s} &=  - \mathcal{B} \overline {I}_t + Z_l^{*-1}\overline V + I_{l}^*+[\overline{V}]^{-1}P_{l}^*. \label{eq:ssIs}
	\end{align}
	\end{subequations}
We can now define the control objective concerning the steady state value of the network voltage, i.e.,
\begin{objective} \emph{(Voltage regulation).}\label{obj:Volt_regulation}
	\begin{align}
	\lim\limits_{t\rightarrow \infty}V_i(t)=V_i^*,~~~~ \forall i \in \mathcal{V},
	\end{align}
	where $V_i^*\in \mathbb{R}_{>0}$ is the desired voltage value of node $i \in \mathcal{V}$.
\end{objective}
In DC networks, the value of the load parameters (i.e., $Z^{*}_l,~I_l^*$ and $P_l^*$) are usually not known. 
As a consequence, it is desired to design a control scheme that does not require the information of the load parameters to achieve Objective \ref{obj:Volt_regulation}. 
In Section \ref{sec:pbc}, we present a passivity-based robust controller that requires (locally) the following system information.
\begin{assumption}\emph{(Available information).}\label{ass:power_load}
$R_{si}$, $L_{si}$ and  $\pi_i \in \mathbb{R}_{\geq 0}$, satisfying $P_{li}^\ast\leq \pi_i$, are available at node $i\in\mathcal{V}$. 
\end{assumption}
	For the sake of convenience, let $\Pi:=\diag \{\pi_1, \cdots, \pi_n\}$. Then, the following inequality holds:
\begin{align}\label{Pi}
[P_l^\ast]\leq \Pi.
\end{align}
\begin{remark}\emph{(Plausibility of Assumption~\ref{ass:power_load}).}
\label{rm:practical_considerations_Pi}
Note that $R_{si}$ and $L_{si}$ are typically components of the converter filter (see Subsection~\ref{subsec:model}) and, consequently, design parameters\footnote{Design-oriented criteria are for instance developed in~\cite{belkhayat1995large} for guaranteeing the stability of electric circuits including P-loads.} whose values are generally known. 
Additionally, Assumption~\ref{ass:power_load} requires the knowledge of an upperbound of the power absorbed by the $i$-th P-load (i.e., $P_{li}^\ast$). 
We notice that in practical applications the power absorbed by any load cannot be infinite and, as a consequence, $\Pi$ in \eqref{Pi} always exists.
Then, the value of $\pi_i$ is generally determined by data analysis and engineering understanding\footnote{Usually, energy end-users (e.g., industrial, commercial and residential applications) have an agreement on the maximum power that they can absorb from the network.}. Moreover, we show in Subsection~\ref{subsec:passivity} and Section~\ref{sec:pbc} that $\pi_i$ is a design parameter for the $i$-th controller and, therefore, its value can be conservatively selected in practice.
For all these reasons, Assumption~\ref{ass:power_load} is reasonable and not restrictive. 
\end{remark}
 
\begin{remark}\emph{(Technical considerations).}
\label{rm:technical_considerations_Pi}
Note that Assumption \ref{ass:power_load} requires that the values of $R_{si}$, $L_{si}$ and $\pi_i$ are available only (locally) at node $i\in\mathcal{V}$. Therefore, it is possible to design decentralized controllers that do not need to exchange information over a communication network and do not depend on the knowledge of the whole network, making the control synthesis simple and the overall control scheme scalable. 
Moreover, Assumption \ref{ass:power_load} does not require any information about the  lines and ZIP-loads parameters, except for an upperbound of only the P component (see Remark~\ref{rm:practical_considerations_Pi}).  
Furthermore, in Subsection~\ref{subsec:exist_passivity_properties} we show that in most of the existing works in the literature, restrictive assumptions on $Z_{li}^\ast, P_{li}^\ast$ and $V_i^\ast$ are generally needed for all $i\in\mathcal{V}$.
\end{remark}
%
%
%
%
%======================================== Passivity
    
    \section{Passivity properties}\label{sec:passivity}
    In this section we first briefly present and discuss some of the existing passivity properties for DC networks including ZIP-loads. 
     Secondly, we establish a novel passivity property for system~\eqref{eq:plant_mg}, leading to the design in Section~\ref{sec:pbc} of a robust passivity-based controller achieving Objective~\ref{obj:Volt_regulation}.

%========================================

    \subsection{Existing passivity properties}\label{subsec:exist_passivity_properties}
There exists a vast amount of literature on possible passive maps for RLC networks (see for instance~\cite{1230225, 1235380,DBLP:journals/corr/abs-1802-02483,253266,machado2018active} and the references therein). In this subsection, we present some of these passivity properties and show their limitations and drawbacks. More precisely, differently from Assumption \ref{ass:power_load}, to the best of our knowledge, the existing passivity properties  require that the Z component of the ZIP  load is \emph{strictly} positive (i.e., $Z_{li}^{*-1}> 0$ for all $i\in\mathcal{V}$) and, additionally, restrictive (sufficient) conditions on $Z_{li}^\ast, P_{li}^\ast$ and $V_i^\ast$ are assumed  to be satisfied for all $i\in\mathcal{V}$.

Motivated by the well-known port-Hamiltonian representation of RLC circuits~\cite{eberard2006energy}, we first introduce the total energy stored in the considered DC network \eqref{eq:plant_mg}, i.e., 
    \begin{equation}\label{storage_energy}
    S(I_s,I_t,V)=\dfrac{1}{2}\|I_s\|^2_{L_s}+\dfrac{1}{2}\|I_t\|^2_{L_t}+\dfrac{1}{2}\|V\|^2_{C_s}.
    \end{equation}
    The storage function \eqref{storage_energy} satisfies 
\begin{align}\label{storagedot_energy}
\dot{S}&=-\|I_t\|^2_{R_t}-\|I_s\|^2_{R_s}-\|V\|^2_{Z_l^{*-1}}-\mathrm{1}^\top P_l^{*}\nonumber\\&\hspace{5mm}- V^\top I_l^{*}+u^\top I_s,
\end{align}
 along the solutions to \eqref{eq:plant_mg}.
    For the sake of convenience, we now define the set of all the solution to \eqref{eq:plant_mg} characterized by positive voltages as follows:
    \begin{align}\label{set:Energy}
    \mathcal{X} :=\{(I_s,I_t,V)\in \R^{2n+m}|V_i> 0, ~\forall i\in \mathcal{V}\}.
    \end{align} 
    Then, the following result holds.
    \begin{proposition}\label{corollary1}\emph{(Stored energy).}
    	System \eqref{eq:plant_mg} is passive with respect to the storage function \eqref{storage_energy} and supply rate $u^\top I_s$, for all the trajectories $(I_s,I_t,V)\in \mathcal{X}$.
    \end{proposition}
\begin{remark}\emph{(Dissipation obstacle).}
Note that because of the notorious {\em dissipation obstacle}\footnote{For system with non-zero supply rate at the desired operating point, the controller has to provide unbounded energy to stabilize the system. In the literature, this is usually referred to as {\em dissipation obstacle} or {\em pervasive dissipation}~\cite{915398,van2000l2}.}, the passivity property presented in Proposition~\ref{corollary1} could be not useful for solving stabilization problem at non trivial operating points.
\end{remark}
  
As an alternative to \eqref{storage_energy}, inspired by the BM theory~\cite{brayton1964theorya,brayton1964theoryb}, which we have briefly recalled in Section~\ref{sec:Gen_BM}, the following result follows from~\cite{1230225, 1235380}.
\begin{proposition} \label{prop:P}\emph{(Generalized mixed potential function).}
Assume that $\left(\lambda,M\right)\in \mathbb{R}\times \mathbb{R}^{(2n+m)^2}$ satisfies $\mathcal{P}_A\geq0$ and $\dot{\mathcal{P}}_A\leq u^\top \dot{I}_s$, for all the trajectories $(I_s,I_t,V)\in \mathcal{X}$, where $\mathcal{P}_A$ is given by \eqref{generalized_BMb}, with $\mathcal{P}$ in \eqref{Mixpot} and $\Gamma, F, \mathcal{G}$ in \eqref{BM_mgb}--\eqref{BM_mgd}.
Then, system~\eqref{eq:plant_mg} is passive with respect to the storage function  $\mathcal{P}_A$ and supply rate $u^\top \dot{I}_s$, for all the trajectories $(I_s,I_t,V)\in \mathcal{X}$.
\end{proposition}
%    \begin{align}\label{storage_genmixpot}
%  S_{\mathcal{P}}(x)=\lambda \mathcal P+\dfrac{1}{2}\|\nabla_x\mathcal P\|_{M}^2,
%  \end{align}
%  where $\mathcal{P}$ is the mixed potential function and $\left(\lambda,M\right)\in \mathbb{R}\times \mathbb{R}^{(2n+m)^2}$ are chosen such that $S_{\mathcal{P}}\geq 0$ and $\dot{S}_{\mathcal{P}}\leq u^\top \dot{I}_s$. 
\begin{remark}\emph{(Choice of $\lambda$ and $M$).}
Note that finding the pair $\left(\lambda,M\right)$ for the considered network \eqref{eq:plant_mg} including ZIP-loads~\eqref{eq:plant_i3} requires a nontrivial endeavor. 
\end{remark}

  Alternatively, in~\cite{DePersis2016,DBLP:journals/corr/abs-1802-02483,253266}, the authors propose the Bregman distance associated to the total energy \eqref{storage_energy} as storage function, i.e.,
  \begin{align}\label{storage_bregman}
  	S_{\mathrm{B}}(I_s,I_t,V)=&~\dfrac{1}{2}\|I_s-\overline{I}_s\|^2_{L_s}+\dfrac{1}{2}\|I_t-\overline{I}_t\|^2_{L_t}\nonumber\\&+\dfrac{1}{2}\|V-{{V^\ast}}\|^2_{C_s}.
  \end{align}
   Along the solutions to \eqref{eq:plant_mg}, the storage function \eqref{storage_bregman} satisfies 
   \begin{align}\label{storagedot_bregman}
   \dot{S}_{\mathrm{B}}=&-\|I_t-\overline{I}_t\|^2_{R_t}-\|I_s-\overline{I}_s\|^2_{R_s}-\|V-{{V^\ast}}\|^2_{G_{\mathrm{B}}(V)}\nonumber\\&+(u-\overline{u})^\top (I_s-\overline{I}_s),
   \end{align}
  where $G_{\mathrm{B}}(V):=Z_l^{*-1}-[P_l^*][V]^{-1}[V^{*}]^{-1}$ represents the equivalent conductance. Define now the following set: 
  \begin{align}\label{set:Bregman}
  \mathcal{X}_{\mathrm{B}} :=\{(I_s,I_t,V) \in \mathcal{X} | G_{\mathrm{B}}(V)\geq 0\}.
  \end{align}
   Then, the following result holds.
    \begin{proposition}\label{corr:shifted}\emph{(Bregman distance).}
  	System \eqref{eq:plant_mg} is (shifted) passive\footnote{{We refer to~\cite{JAYAWARDHANA2007618,MONSHIZADEH201955} for further details on shifted passivity}.} with respect to the storage function \eqref{storage_bregman} and supply rate $(u-{u^\ast})^\top (I_s-\overline{I}_s)$, for all the trajectories $(I_s,I_t,V)\in \mathcal{X}_{\mathrm{B}}$.
  \end{proposition}
  \begin{remark}\label{rm:restrictive_conditions}\emph{(Restrictive conditions on $Z^{*}_{li}, P_{li}^*$ and $V_i^*$).}
From \eqref{set:Bregman} it straightforwardly follows that the absence of constant impedance loads makes the set $\mathcal{X}_{\mathrm B}$ empty and, consequently, the sufficient condition $(I_s,I_t,V)\in \mathcal{X}_{\mathrm{B}}$ in Corollary~\ref{corr:shifted} is not satisfied. More precisely, we notice that $\mathcal{X}_{\mathrm B}$ is empty if there exists at least a node, say node $j$, where $Z^{*-1}_{lj}=0, j\in \mathcal{V}$.
Moreover, the voltage reference $V^*_i$ must satisfy the inequality $V^*_i\geq \sqrt{Z_{li}^*P_{li}^*}$, for all $i\in\mathcal{V}$. However, we observe that in practical applications, the values of the load parameters are generally unknown and, even if they are estimated, the voltage reference cannot be chosen arbitrarily large. For all these reasons, in practical applications $\mathcal{X}_{\mathrm B}$ could be often empty and/or not contain the steady-state solution corresponding to the voltage reference $V^\ast_i, i\in\mathcal{V}$.
  \end{remark}
More recently, for the sake of robustness, in~\cite{2018arXiv181102838C,Cucuzzella_arxiv2019,NOLCOS}, the authors have proposed the following Krasovskii's Lyapunov function as storage function:
   \begin{eqnarray}\label{storage_krasovskii}
S_{\mathrm{K}}(I_s,I_t,V,u)=\dfrac{1}{2}\|\dot{I}_s\|^2_{L_s}+\dfrac{1}{2}\|\dot{I}_t\|^2_{L_t}+\dfrac{1}{2}\|\dot{V}\|^2_{C_s},
\end{eqnarray}
which satisfies, along the solutions to \eqref{eq:plant_mg}, 
\begin{align}\label{storagedot_krasovskii}
\dot{S}_{\mathrm{K}}=&-\|\dot{I}_t\|^2_{R_t}-\|\dot{I}_s\|^2_{R_s}-\|\dot{V}\|^2_{G_{\mathrm{K}}(V)}+\dot{u}^\top \dot{I}_s,
\end{align}
where $G_{\mathrm{K}}(V):={Z_l^{*-1}-[P_l^*][V]^{-2}}$ represents the equivalent conductance.  Define now the following set: 
\begin{align}\label{set:Krasovskii}
\mathcal{X}_{\mathrm{K}} :=\{(I_s,I_t,V,u)\in\mathcal{X}\times\R^{n}_{>0}|G_{\mathrm{K}}(V)\geq 0\}.
\end{align}
   Then, the following result holds.
\begin{proposition}\label{corr:kras}\emph{(Krasovskii's Lyapunov function).}
	System \eqref{eq:plant_mg} is passive with respect to the storage function \eqref{storage_krasovskii} and supply rate $\dot{u}^\top \dot{I}_s$, for all the trajectories $(I_s,I_t,V,u)\in\mathcal{X}_{\mathrm{K}}$.
\end{proposition}

\begin{remark}\label{rm:restrictive_conditions2}\emph{(Restrictive conditions on $Z^{*}_{li}, P_{li}^*$ and $V_i^*$).}
In order for $\mathcal{X}_{\mathrm{K}}$ to be nonempty and contain the steady-state solution corresponding to the voltage reference $V^\ast_i, i\in\mathcal{V}$, restrictive (sufficient) conditions (similar to those discussed in Remark~\ref{rm:restrictive_conditions}) on $Z^{*}_{li}, P_{li}^*$ and $V_i^*$ need to be satisfied for all $i\in\mathcal{V}$.
\end{remark}

In the next section, we propose a \emph{novel} passifying input and a storage function based on the generalized mixed potential function \eqref{generalized_BMb}, leading to a passivity property  for every type of load (even loads consisting of only the P component), for every positive voltage reference and for all the trajectories evolving in the subspace of the state-space where the voltage is positive, i.e., for every $(I_s,I_t,V)\in \mathcal{X}$, with $\mathcal{X}$ defined in~\eqref{set:Energy}.

%========================================

\subsection{A novel passifying input}\label{subsec:passivity}
In the previous section, we have shown that the passive output of the existing passivity properties for general RLC circuits is the current or its time derivative. 
However, from the dissipation inequalities %\eqref{storagedot_energy},
\eqref{storagedot_bregman} and \eqref{storagedot_krasovskii}, we can observe that in order to counteract the effects of the P-loads, it would be desired that the passive output is (function of) the voltage or its time derivative, allowing for the injection of extra damping into the controlled system. 
In the following theorem, inspired by the generalized BM structure presented in Proposition~\ref{lemma:gradient_generalized}, we establish a new passivity property with output port-variable equal to the first time derivative of the voltage. This property is essential in Section~\ref{sec:pbc} to design the proposed robust passivity-based control achieving Objective~\ref{obj:Volt_regulation}.
\begin{theorem}\emph{(Novel passivity property).}\label{thm:proposed_passivity}
	Let Assumption~\ref{ass:power_load} hold. Given system \eqref{eq:plant_mg}, define the mapping $u_{\mathrm{PBC}}:\mathcal{X}\rightarrow \mathbb{R}^n$ as follows:
		\begin{align}
		u_{\mathrm{PBC}}&:=R_sI_s+Q_0(x)\dot{x},\label{PBC_BM}
		\end{align}
	where $Q_0:\mathcal{X} \rightarrow \mathbb{R}^{n\times (2n+m)}$. Consider the following mixed potential function:
	\begin{equation}
	\label{eq:new_mixed_potential_func}
	\mathcal{P}(I,V)=\dfrac{1}{2}\|I_t\|_{R_t}^2+I^\top\Gamma V-\mathcal{G}(V),
	\end{equation}
	where $\Gamma$ and $\mathcal{G}(V)$ are given by \eqref{BM_mgb} and \eqref{BM_mgd}, respectively.
	 Then, the following statements hold:
	\begin{enumerate}
		\item[(i)] Let $\upsilon\in \mathbb{R}^n$. Consider the following input:
		\begin{align}\label{PBC_BM_1}
			u=u_{\mathrm{PBC}}+L_s\upsilon.
		\end{align}
		The closed-loop system \eqref{eq:plant_mg}, \eqref{PBC_BM_1} is described by the generalized gradient structure \eqref{gradient_generalized} with input $\upsilon$ and mixed potential function \eqref{eq:new_mixed_potential_func}.
	\item[(ii)] Let  
	 $Q_0=\begin{bmatrix}
	\mathcal{O}_{n}&\mathcal{O}_{n\times m}&-L_s\Pi[V]^{-2}
	\end{bmatrix}$. 
	The closed-loop system \eqref{eq:plant_mg}, \eqref{PBC_BM_1} is passive with respect to the storage function $\mathcal{P}_A$ in \eqref{generalized_BMb} and supply rate $\upsilon^\top \dot{V}$, for all the trajectories $(I_s,I_t,V)\in \mathcal{X}$.
	\end{enumerate}
\end{theorem}
\begin{pf}
Part \emph{(i)}. Along the solutions to \eqref{eq:plant_mg}, $\mathcal{P}_A$ in \eqref{generalized_BMb} (with $\mathcal P$ given by \eqref{eq:new_mixed_potential_func}) satisfies
	\begin{align}\label{gradient_clp_mg}
	\nabla_x\mathcal{P}_A&=(\lambda \mathcal{I}_{2n+m}+\nabla_x^2\mathcal{P}M)\nabla_x\mathcal{P}\nonumber\\
%	&=(\lambda \mathcal{I}_{2n+m}+\nabla_x^2\mathcal{P}M)\begin{bmatrix}
%	V \\
%	R_t I_t +\mathcal{B}^\top V\\
%	I_{s} + \mathcal{B}I_t - I_l(V)	
%	\end{bmatrix}\nonumber\\
	&=(\lambda \mathcal{I}_{2n+m}+\nabla_x^2\mathcal{P}M)\begin{bmatrix}
	V+R_sI_s-R_sI_s \\
	R_t I_t +\mathcal{B}^\top V\\
	I_{s} + \mathcal{B}I_t - I_l(V)	
	\end{bmatrix}\nonumber\\
%	&= (\lambda \mathcal{I}_{2n+m}+\nabla_x^2\mathcal{P}M)(Q\dot{x}-\tilde{B}u+\tilde{B}R_sI_s)\nonumber\\
	&= (\lambda \mathcal{I}_{2n+m}+\nabla_x^2\mathcal{P}M)(Q\dot{x}-\tilde{B}(-R_sI_s+u))\nonumber\\
%	&= (\lambda \mathcal{I}_{2n+m}+\nabla_x^2\mathcal{P}M)(Q\dot{x}-\tilde{B}(Q_0\dot{x}+L_s\upsilon))\nonumber\\
	&= (\lambda \mathcal{I}_{2n+m}+\nabla_x^2\mathcal{P}M)((Q-\tilde{B}Q_0)\dot{x}-\tilde{B}L_s\upsilon)\nonumber\\
	&= Q_A\dot{x}-\tilde{B}_A\upsilon,
	\end{align}
 where in \eqref{generalized_BMc} and \eqref{generalized_BMd} we use $D=L_s$. Part  \emph{(ii)}. Let $\lambda=0$ and $M=\diag\{L^{-1}_s,L^{-1}_t,C^{-1}_s\}$. Then, the generalized mixed potential function $\mathcal{P}_A$ in \eqref{generalized_BMb} (with $\mathcal P$ given by \eqref{eq:new_mixed_potential_func}) can be expressed as follows:
 \begin{align}\label{storage_mg}
					\mathcal{P}_A(I_s,I_t,V) = &~\dfrac{1}{2} \|V\|_{L_s^{-1}}^2+\dfrac{1}{2}\|R_t I_t +\mathcal{B}^\top V\|_{L_t^{-1}}^2\nonumber\\&+\dfrac{1}{2}\|I_{s} + \mathcal{B}I_t - I_l(V)\|_{C_s^{-1}}^2.
				\end{align}
  Furthermore, $\tilde{B}_A=[\mathcal{O}_{n\times (n+m)}~-\mathcal{I}_n]^\top $ and $Q_A$ in \eqref{generalized_BMa} can be expressed as follows:
	\begin{equation*}
		Q_A=%&~\nabla_x^2\mathcal{P}M(Q-\tilde{B}Q_0)\\
%		&=\begin{bmatrix}
%		\mathcal{O}_{n}&\mathcal{O}_{n\times m}&\mathcal{I}_n\\\mathcal{O}_{m\times n}&R_t&\mathcal{B}^\top\\\mathcal{I}_n&\mathcal{B}&-G_{\mathrm{K}}(V)%Z_l^{*-1}+[P_l^*][V]^{-2}
%		\end{bmatrix}\begin{bmatrix}
%		-\mathcal{I}_n&\mathcal{O}_{n\times m}&-\Pi[V]^{-2}\\\mathcal{O}_{m\times n}&-\mathcal{I}_m&\mathcal{O}_{m\times n}\\\mathcal{O}_{n}&\mathcal{O}_{n\times m}&\mathcal{I}_n
%		\end{bmatrix}\\
		\begin{bmatrix}
		\mathcal{O}_{n}&\mathcal{O}_{n\times m}&\mathcal{I}_n\\\mathcal{O}_{m\times n}&-R_t&\mathcal{B}^\top\\-\mathcal{I}_n&-\mathcal{B}&-G_{\Pi}(V)
		\end{bmatrix},
	\end{equation*}
	where, for every $(I_s,I_t,V)\in \mathcal{X}$, $G_{\Pi}:=Z_l^{*-1}+(\Pi-[P_l^*])[V]^{-2}$ represents the equivalent conductance and is positive semi-definite by virtue of Assumption~\ref{ass:power_load}.
	Therefore, $Q_A+Q_A^\top  \leq 0$. 
	Moreover, the generalized mixed potential function~\eqref{storage_mg} satisfies 
\begin{equation}\label{eq:S_dot}
		\dot{\mathcal{P}}_A%&=\nabla_x^\top \mathcal{P}_A\dot{x}\nonumber\\
		%&=\left(Q_A\dot{x}-\tilde{B}_A\upsilon\right)^\top \dot{x}\nonumber\\
		%&=\frac{1}{2}\dot{x}\left(Q_A+Q_A^\top\right)\dot{x}-\upsilon^\top \tilde{B}^\top _A\dot{x}\nonumber\\
		= -\|\dot{I}_t\|_{R_t}^2-\|\dot{V}\|_{G_{\Pi}(V)}^2+\upsilon^\top \dot{V},
\end{equation}
along the solutions to \eqref{gradient_clp_mg}, implying that the closed-loop system is passive with port-variables $\upsilon$ and $\dot{V}$.
$\hfill\blacksquare$
\end{pf}
\begin{figure}[t]
\centering
\includegraphics[width=\columnwidth]{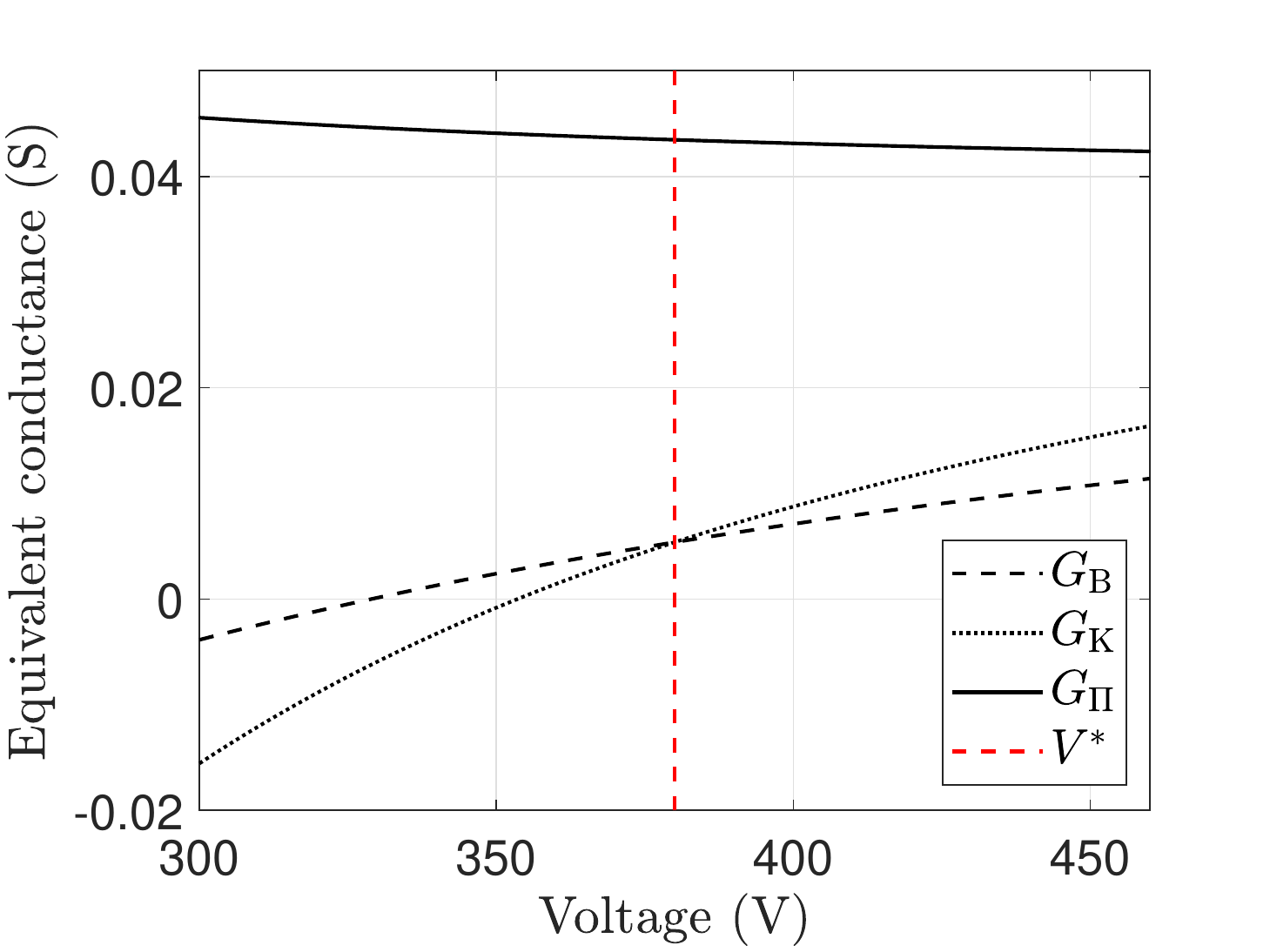}
	\caption{Comparison between $G_{\mathrm{B}}(V)$,  $G_{\mathrm{K}}(V)$ and  $G_{\Pi}(V)$. 
	}
	\label{fig:Impedance}
	\vspace{.2cm}
\end{figure}

\begin{remark}\emph{(Novel passifying input).}
	The input $u_{\mathrm{PBC}}$ in \eqref{PBC_BM} with $Q_0=\begin{bmatrix}
	0&0&-L_s\Pi[V]^{-2}
	\end{bmatrix}$ becomes
	\begin{equation}\label{eq:stab_cont_mg1}
	u_{\mathrm{PBC}}=R_sI_s-L_s\Pi[V]^{-2}\dot{V}.
	\end{equation}
	It is worth noting that the input \eqref{eq:stab_cont_mg1} has a decentralized structure and the time derivative of the voltage can be robustly estimated in a finite time by implementing for instance the well-known Levant's differentiator~\cite{Levant03}. Moreover, the term $-L_s\Pi[V]^{-2}$ in \eqref{eq:stab_cont_mg1} can be interpreted as the adaptive gain of a proportional action on the passive output $\dot V$ (or, equivalently, the adaptive gain of a derivative action on the voltage error $V-V^\ast$). 
	%\red{A common step in this methodology is to cancel the effect of series filter resistance, called as Negative Input Resistance Compensation (NIRC) \cite{4380518}.}
\end{remark}

For the sake of exposition, consider as illustrative example a ZP-load with $Z_l^{*-1}=$ \SI{0.04}{\siemens}, $P_l^*=$ \SI{5e3}{\watt} and nominal voltage $V^\ast=$ \SI{380}{\volt}. Figure \ref{fig:Impedance} shows the comparison between the equivalent conductances $G_{\mathrm{B}}(V)$,  $G_{\mathrm{K}}(V)$ and  $G_{\Pi}(V)$, with $\Pi=$ \SI{5.5e3}{\watt}. Differently from $G_{\mathrm{B}}(V)$ and $G_{\mathrm{K}}(V)$, which are positive only around the desired operating point, i.e.,  $V=V^\ast=$ \SI{380}{\volt}, $G_\Pi(V)$ is positive for any $V>0$.
%
%
%
%
%======================================== PBC

\section{Robust Passivity-based control}\label{sec:pbc}
In this section, we use the new passivity property established in Theorem~\ref{thm:proposed_passivity} to design a controller that stabilizes the closed-loop system and achieves Objective~\ref{obj:Volt_regulation} despite the uncertainty affecting the load components. 
As discussed in the previous section, usually, the output port-variable is (function of) the current or its first time derivative. Therefore, current controllers are often designed. However, evaluating the value of the current corresponding to the desired voltage reference, generally requires the complete knowledge of the load, which we reasonably suppose to be uncertain (see Assumption~\ref{ass:power_load}). 
In order to avoid this issue, we have established a passive map with the output port-variable equal to the first time derivative of the voltage. As a consequence, we simply shape the closed-loop storage function (introducing a function of the voltage) such that it has a minimum at the desired operating point. More precisely, we adopt the {\em output-shaping} technique introduced in~\cite{2018arXiv181102838C}, where the integrated passive output is used to shape the closed-loop storage function. 
\begin{theorem}\emph{(Closed-loop stability).}\label{prop::stab_ZIP_buck_mg}
	Let Assumption \ref{ass:power_load} hold.
	Given system \eqref{eq:plant_mg}, define the mapping $u_{\mathrm{Stab}}:\mathcal{X}\rightarrow \mathbb{R}^n$ as follows:
	\begin{align}\label{eq:stab_cont_mg2}
\begin{split}
u_{\mathrm{Stab}} :=& -L_sK_1(V-V^*)-L_sK_2\dot{V}+V^{*},
\end{split}
\end{align}	
$K_1\geq 0,K_2> 0 \in \mathbb{R}^{n\times n}$, $V^\ast\in\R^n_{>0}$.  
	Then, the following statements hold:
	\begin{enumerate}
		\item [(i)] Let $\mu\in\R^{n}$. Consider the following input:
		\begin{equation}\label{eq:u_cont_stab_mg}
		u=L_s\mu+u_{\mathrm{PBC}}+u_{\mathrm{Stab}},
		\end{equation}
		with  $u_{\mathrm{PBC}}$ and $u_{\mathrm{Stab}}$ given by \eqref{eq:stab_cont_mg1} and \eqref{eq:stab_cont_mg2}, respectively.
		The closed-loop system \eqref{eq:plant_mg}, \eqref{eq:u_cont_stab_mg} is passive with respect to the supply rate $\mu^\top\dot{V}$ and storage function $S_d=\mathcal{P}_A+S_a$, with $\mathcal{P}_A$ in \eqref{storage_mg} and $S_a$ defined as follows:
\begin{equation}\label{eq:storage_fun_Sa_mg}
S_a(V) := \dfrac{1}{2}\|V-V^{*}\|_{K_1}^2-V^\top L^{-1}_sV^{*}+\dfrac{1}{2}\|V^*\|_{L_s^{-1}}^2.
\end{equation}
		\item [(ii)] Consider the closed-loop system~\eqref{eq:plant_mg}, \eqref{eq:u_cont_stab_mg} with $\mu$ equal to zero. Then, the equilibrium $(\overline{I}_s, \overline{I}_t, V^{*})\in\mathcal{X}$ is asymptotically stable in $\mathcal{X}$, with $\mathcal{X}$ defined in \eqref{set:Energy}.
	\end{enumerate}
\end{theorem}
\begin{pf}
Part \emph{(i)}. Consider the storage function $\mathcal{P}_A$ in \eqref{storage_mg}. Now, we use a function of the integrated output port-variable, i.e., $S_a:\R^n_{>0}\rightarrow\R$ defined in \eqref{eq:storage_fun_Sa_mg}, to shape the desired closed-loop storage function $S_d:\mathcal{X}\rightarrow\R$. More precisely, $S_d=\mathcal{P}_A+S_a$ can be expressed as follows:
\begin{align}\label{eq:clp_storage}
\begin{split}
S_d=&~\dfrac{1}{2}\|V-V^{*}\|_{L_s^{-1}+K_1}^2+\dfrac{1}{2}\|R_t I_t +\mathcal{B}^\top V\|_{L_t^{-1}}^2\\
&+\dfrac{1}{2}\|I_{s} + \mathcal{B}I_t - I_l(V)\|_{C_s^{-1}}^2.
\end{split}
\end{align}
Moreover, along the closed-loop dynamics \eqref{eq:plant_mg},  \eqref{eq:u_cont_stab_mg}, $S_a$ satisfies $\dot{S}_a=\dot{V}^\top \left(K_1(V-V^*)-L_s^{-1}V^*\right)$, and $S_d$ satisfies
	\begin{subequations}\label{sdot_closed_loop}
			\begin{align}
		\dot{S}_d&=\dot{\mathcal{P}}_A+\dot{S}_a\label{sdot_closed_loopa}\\
%		&=-\|\dot{I}_t\|_{R_t}^2-\|\dot{V}\|_{G_{\Pi}(V)}^2+\upsilon^\top \dot{V}+\dot{S}_a\label{sdot_closed_loopb}\\
		&=-\|\dot{I}_t\|_{R_t}^2-\|\dot{V}\|_{G_{\Pi}(V)}^2+\mu^\top \dot{V}\nonumber\\&\hspace{5mm}+\dot{V}^\top \left(-K_1(V-V^*)-K_2\dot{V}+L_s^{-1}V^*\right)\nonumber\\&\hspace{5mm}+\dot{V}^\top \left(K_1(V-V^*)-L_s^{-1}V^*\right)\label{sdot_closed_loopc}\\
		&=-\|\dot{I}_t\|_{R_t}^2-\|\dot{V}\|_{G_{\Pi}(V)+K_2}^2+\mu^\top \dot{V},\label{sdot_closed_loopd}
		\end{align}
	\end{subequations}
with $\dot{\mathcal{P}}_A$ given by \eqref{eq:S_dot}, $\upsilon=\mu+L_s^{-1}u_{\mathrm{Stab}}$ and $u_{\mathrm{Stab}}$ given by~\eqref{eq:stab_cont_mg2}. Part~\emph{(ii)}. System \eqref{eq:plant_mg} in closed-loop with $u=u_{\mathrm{PBC}}+u_{\mathrm{Stab}}$ becomes
	\begin{align}\label{eq:V_inter_S_d_mg}
	\begin{split}
	\dot{I}_s+\left(\Pi[V]^{-2}+K_2\right)\dot{V}&=-\left(K_1+L_s^{-1}\right)(V-V^*),\\
	-	L_t\dot{I}_t & = R_t I_t +\mathcal{B}^\top V,\\
	C_{s}\dot{V} & = I_{s} + \mathcal{B}I_t - I_l(V).		
	\end{split}
		\end{align}
Now, we observe that $S_d$ in \eqref{eq:clp_storage} is positive and attains a minimum at the (unique) equilibrium point $(\overline{I}_s,\overline{I}_t,V^*)\in\mathcal{X}$ of the closed-loop system \eqref{eq:V_inter_S_d_mg}. Then, we use $S_d$ in  \eqref{eq:clp_storage} as a candidate Lyapunov function. Therefore, \eqref{sdot_closed_loopd} implies that there exists a forward invariant set $\Upsilon$ and by Lasalle's invariance principle the solutions that start in $\Upsilon$ approach to the largest invariant set contained in 
	\begin{equation}\label{set:forward_inv_set_thm1_mg}
	\Upsilon \cap \left\{\left(I_s,I_t,V\right)\in \mathcal{X}:\dot{V}=0,~\dot{I}_t=0 \right\}.
	\end{equation}
	On this invariant set, by differentiating the third line of \eqref{eq:V_inter_S_d_mg}, it follows that $\dot{I}_s=0$. Moreover, from \eqref{eq:V_inter_S_d_mg},  it also follows that $V=V^{*}$, $I_s=\overline{I}_s$ and $I_t=\overline{I}_t$. 
	$\hfill\blacksquare$
	\end{pf}
\begin{remark}\emph{(Control law).}\label{rem:controller}
	The control law \eqref{eq:u_cont_stab_mg} with $\mu=0$ can be written compactly as follows:
\begin{equation}\label{eq:stab_cont}
u=\underbrace{R_sI_s+V^{*}}_{\text{\emph{(a)}}}-\underbrace{L_sK_1(V-V^*)}_{\text{\emph{(b)}}}-\underbrace{L_s(\Pi[V]^{-2}+K_2)\dot{V}}_{\text{\emph{(c)}}}.
\end{equation}
In \eqref{eq:stab_cont}, the term \emph{(a)} simply represents a feedforward control action. The term \emph{(b)} represents an integral action on the passive output (or, equivalently, a proportional action on the voltage error), with gain $-L_sK_1$. Finally, the term \emph{(c)} represents a proportional action on the passive output (or, equivalently, a derivative action on the voltage error), with adaptive gain $-L_s(\Pi[V]^{-2}+K_2)$. Moreover, the voltage dynamics of node $i \in \mathcal{V}$ can be expressed as
\begin{equation*}
\ddot{V}_i + \frac{\frac{1}{Z_{li}^\ast}-\frac{P_{li}^\ast}{V_i^2}+\frac{\pi_i}{V_i^2}+K_{2i}}{C_{si}}\dot{V}_i + \frac{K_{1i}L_{si}+1}{L_{si}C_{si}}V_i = \frac{V_i^\ast}{L_{si}C_{si}},
\end{equation*}
implying that a nonlinear virtual resistor (with conductance ${\pi_i}/{V_i^2}+K_{2i}$) is connected in parallel to the real $i$-th P-load (see~\cite[Subsection V.A]{1323174} for further details on parallel damping injection).
\end{remark}

\begin{remark}\emph{(Robustness).}\label{rem:robustness}
 Note that the controller \eqref{eq:stab_cont} does not require any information about the  lines and load parameters $Z_l^{*-1}$, $I_l^*$ and $P_l^*$, except for an upperbound of only the P component (see Remark~\ref{rm:practical_considerations_Pi}). However, the proposed controller is not robust with respect to uncertainty affecting the filter impedance $R_s L_s$ (see Remark~\ref{rm:practical_considerations_Pi}).
\end{remark}
%
%
%
%
%======================================== Sim

\section{Numerical results}\label{sec:sim}
In this section, first we present an illustrative example to show the benefits of the passivity property established in Subsection~\ref{subsec:passivity} with respect to the shifted and Krasovskii's passivity properties discussed in Subsection~\ref{subsec:exist_passivity_properties}. Secondly, we show the performance of the proposed control scheme on a DC network comprising four nodes with a ring topology.
%========================================

\subsection{Illustrative example}
\begin{figure*}[t]
\center
\subfigure[case 1: $P_l^\ast=$ \SI{5e3}{\watt}]{
\centering
\includegraphics[width=\columnwidth]{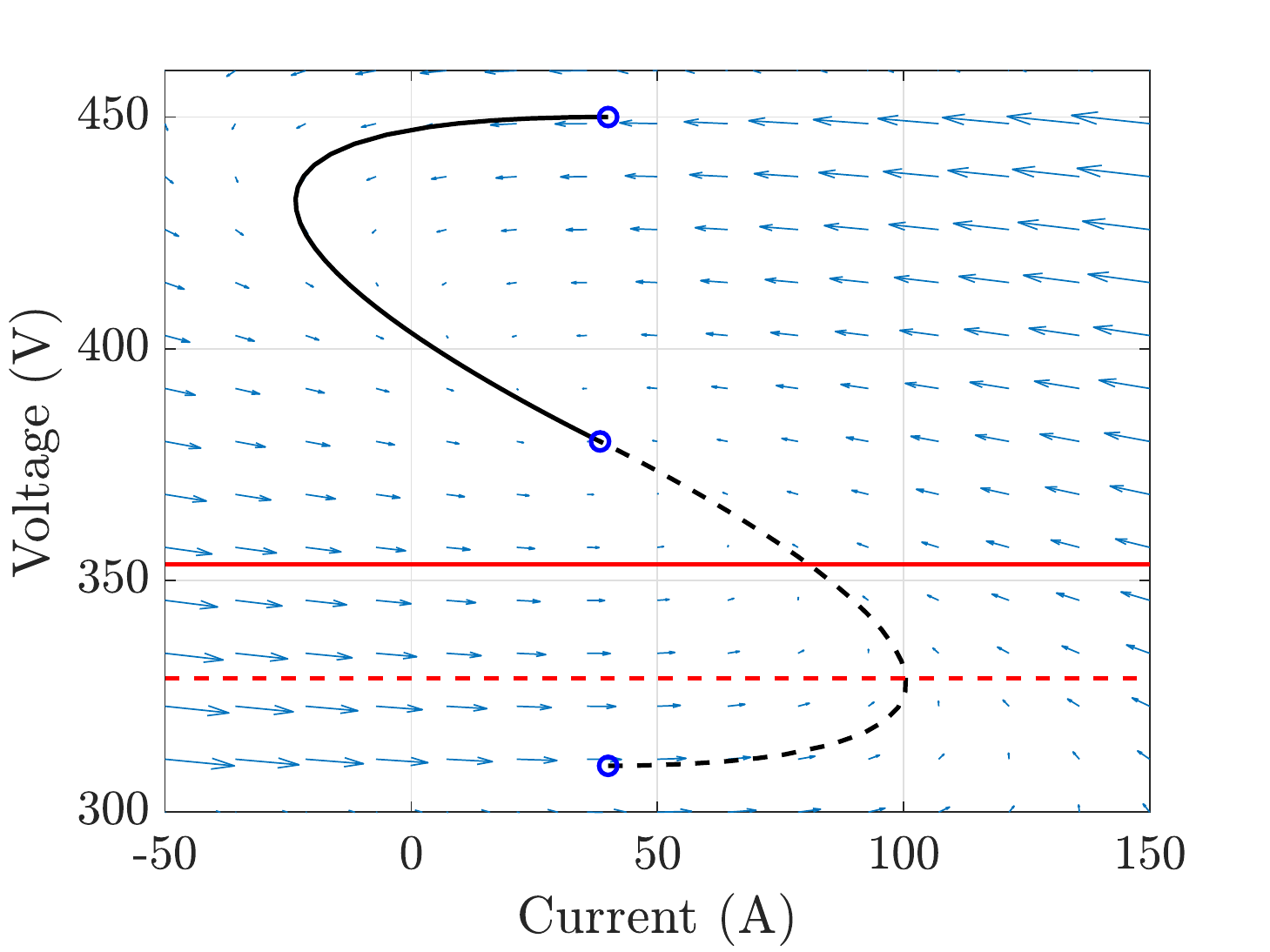}
}
\hspace{0mm}
\subfigure[case 2: $P_l^\ast=$ \SI{6.5e3}{\watt}]{
\centering
\includegraphics[width=\columnwidth]{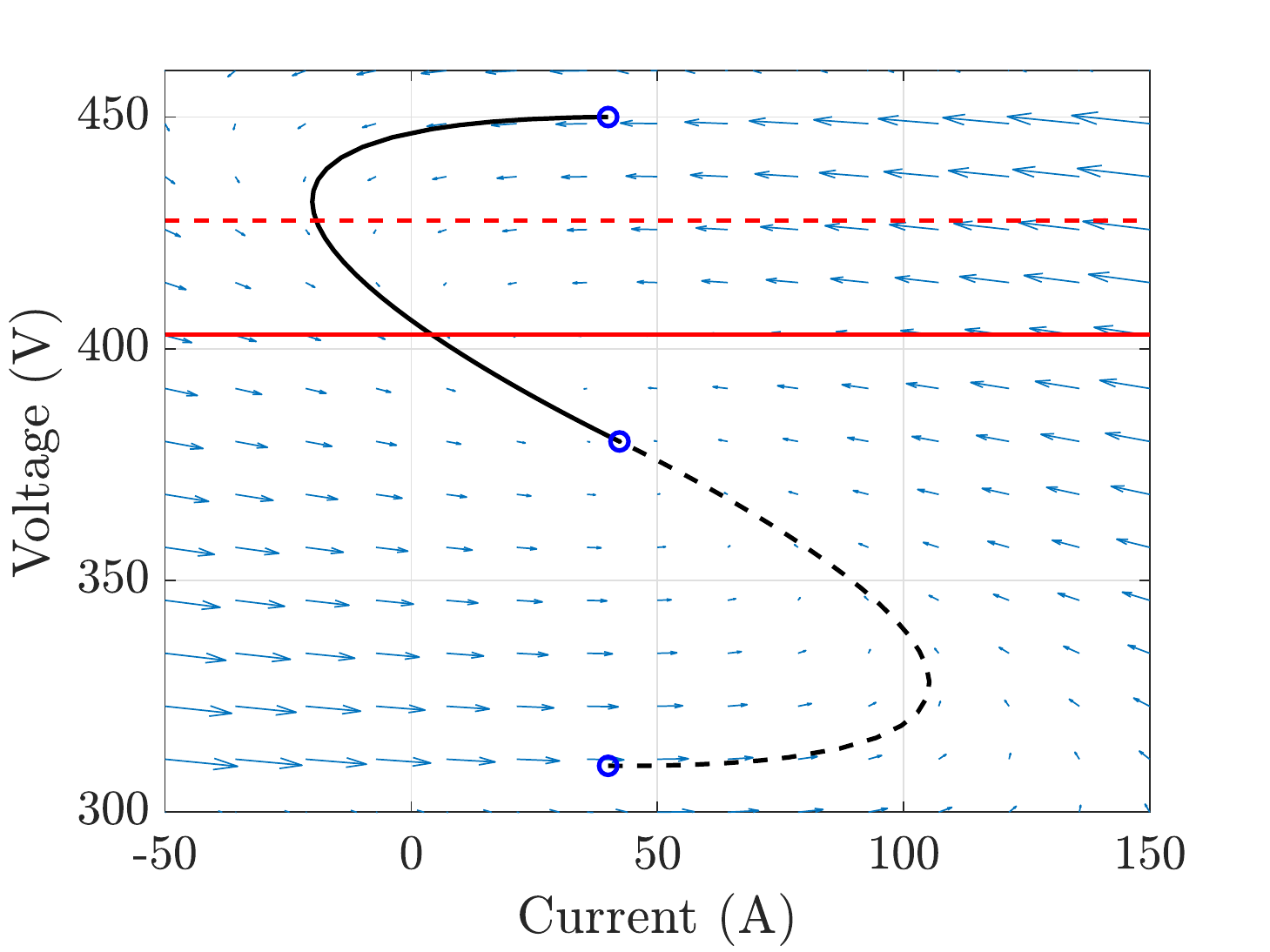}
}
	\caption{Vector field and state-space trajectories of an RLC circuit including a ZIP-load, controlled by \eqref{eq:stab_cont}. The areas above the dashed and solid red lines represent the sets $\mathcal{X}_{\mathrm{B}}$, and $\mathcal{X}_{\mathrm{K}}$, respectively. 
	}
	\label{fig:awesome_image1}
	\vspace{.2cm}
\end{figure*}

\begin{table}[t]
 	 \centering
	 	\caption{P-load and equivalent conductance.} 
	 	\vspace{.2cm}
	 	{\begin{tabular}{c|cc}
case			&1			&2\\			
\hline	 						
	 			$P_l^*$ (W)					&\num{5e3}			&\num{6.5e3}\\			
	 			$Z_l^{*-1}-[P_l^*][V^{*}]^{-2}$ (S)	&\num{0.0054}			&\num{-0.0050}
	 	\end{tabular}}
	 	\label{tab:parameters_remark}
	 	\vspace{.2cm}
	 \end{table}

For the sake of exposition, consider as illustrative example a simple RLC ($R_s=$ \SI{10}{\milli\ohm}, $L_s=$ \SI{1.12}{\milli\henry}, $C_s=$ \SI{6.8}{\milli\farad}) circuit including a ZIP ($Z_l^{\ast-1}=$ \SI{0.04}{\siemens}, $I_l^\ast=$ \SI{10}{\ampere}) load (see the scheme of DGU $i$ in Figure~\ref{fig:networks}). The values assumed by the P component are reported in Table~\ref{tab:parameters_remark} together with the equivalent conductance $G_{\mathrm{B}}(V^*)=G_{\mathrm{K}}(V^*)=Z_l^{*-1}-[P_l^*][V^{*}]^{-2}$, evaluated at the desired voltage reference $V^\ast=$ \SI{380}{\volt}.
	In \eqref{eq:stab_cont}, we select $K_1=$ \num{1}, $K_2=$ \num{5} and $\Pi=$ \SI{10e3}{\watt}. 
	The vector field of the closed-loop system is illustrated in Figure \ref{fig:awesome_image1}, showing that \eqref{eq:stab_cont} is a stabilizing control law. The solid black line represents the state-space trajectory starting from (\SI{40}{\ampere}, \SI{450}{\volt}), while the dashed black line represents the state-space trajectory starting from (\SI{40}{\ampere}, \SI{310}{\volt}). 
	Moreover, Figure~\ref{fig:awesome_image1} shows the sets $\mathcal{X}_{\mathrm{B}}$ (the area above the dashed red line) and $\mathcal{X}_{\mathrm{K}}$ (the area above the solid red line) defined in \eqref{set:Bregman} and \eqref{set:Krasovskii}, respectively.  
	More precisely, when $P_l^*$ is equal to \SI{5e3}{\watt} (see Figure~\ref{fig:awesome_image1}a), the value of the equivalent conductance evaluated at the desired voltage reference is positive (see Table~\ref{tab:parameters_remark}) and, therefore, there exists $(\overline{I}_s,V^*)$ belonging to $\mathcal{X}_{\mathrm{B}}$ and $\mathcal{X}_{\mathrm{K}}$. Indeed, it is straightforward to verify that the desired equilibrium point ($\overline{I}_s=$ \SI{38.36}{\ampere}, $V^\ast=$ \SI{380}{\volt}) is located above the red lines, where $\overline{I}_s$ satisfies~\eqref{eq:ssIs}. 
	However, when $P_l^*$ is equal to \SI{6.5e3}{\watt} (see Figure~\ref{fig:awesome_image1}b), the value of the equivalent conductance evaluated at the desired voltage reference becomes negative (see Table~\ref{tab:parameters_remark}) and, therefore, $\mathcal{X}_{\mathrm{B}}$ and $\mathcal{X}_{\mathrm{K}}$ do not contain the steady-state solution corresponding to the desired voltage reference $V^\ast$. Also in this case, it is indeed straightforward to verify that the desired equilibrium point ($\overline{I}_s=$ \SI{42.31}{\ampere}, $V^\ast=$ \SI{380}{\volt}) is located below the red lines. 
	
We can conclude from this simple example that the (sufficient) conditions imposed on the parameters and trajectories of the system by the shifted and Krasovskii's passivity properties (see Propositions~\ref{corr:shifted} and~\ref{corr:kras} in Subsection~\ref{subsec:exist_passivity_properties}) can be very restrictive. Differently, the proposed passivity-based approach does not require any condition on the system parameters and allows the system trajectories to evolve in the subspace of the state-space where the voltage is (simply) positive. 

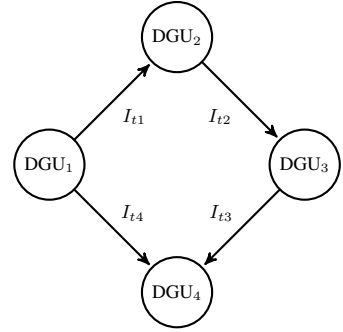
\begin{figure}[t]
\begin{center}
\begin{tikzpicture}[scale=0.8,transform shape,->,>=stealth',shorten >=1pt,auto,node distance=3cm,
                    semithick]
  \tikzstyle{every state}=[circle,thick,fill=white,draw=black,text=black]

  \node[state] (A)                    {DGU$_1$};
  \node[state]         (B) [above right of=A] {DGU$_2$};
  \node[state]         (D) [below right of=A] {DGU$_4$};
  \node[state]         (C) [below right of=B] {DGU$_3$};

  \path (A) edge   [below] node {\hspace{7mm}$I_{t1}$} (B)
  		edge 	     node {$I_{t4}$} (D)
           (B) edge      [below]        node {\hspace{-7mm}$I_{t2}$} (C)
           (C) edge         [above left]     node {$I_{t3}$} (D);

\end{tikzpicture}
\caption{Scheme of the considered DC power network with four nodes and four  lines. The arrows indicate the positive direction of the currents through the power network.}
\label{fig:networks_example}
\end{center}\vspace{.2cm}
\end{figure}

%======================================== 

\subsection{DC network}

The proposed decentralized control scheme is now assessed in simulation, considering a DC network comprising four nodes interconnected as shown in Figure~\ref{fig:networks_example}. 
The control gains in \eqref{eq:stab_cont} are $K_1=$ \num{50} and $K_2=$ \num{200}, and the parameters of each node and line are chosen as in~\cite{7934339} and reported in Tables \ref{tab:parameters1} and~\ref{tab:parameters2}, respectively. For the sake of illustration, we show two different simulation scenarios. 

\begin{figure}[t]
	\centering
	\includegraphics[trim=0cm 0cm 0cm 0cm, clip=true, width=\columnwidth]{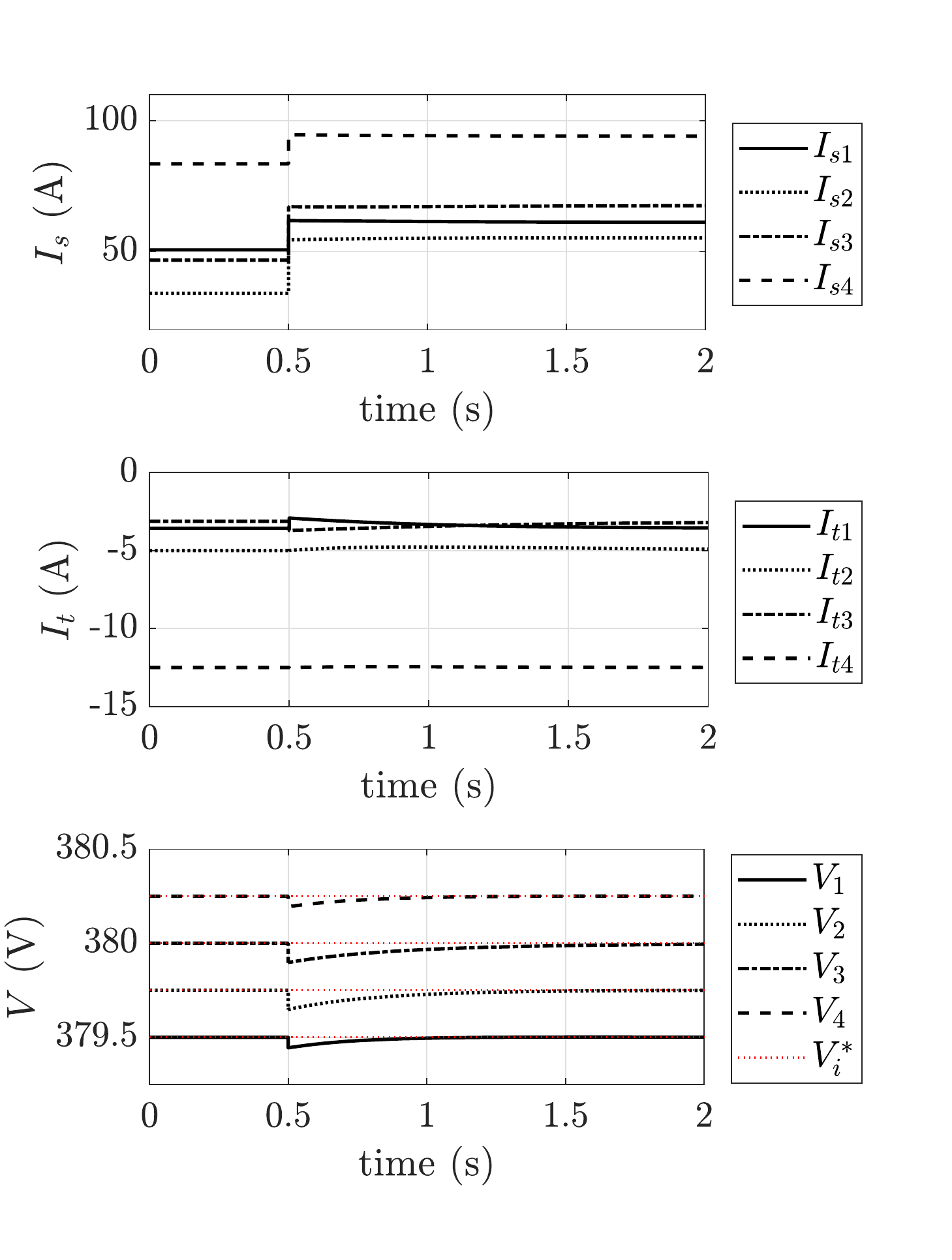}
	\caption{Scenario 1. From the top: time evolution of the generated currents, currents exchanged through the lines, and voltages together with the corresponding references.}
	\label{fig:sen_a}
	\vspace{.2cm}
\end{figure}

{\bf{Scenario  1}} (ZIP-loads). In this scenario we consider that all the loads of the network are ZIP-loads.   
At the time instant $t=$ \SI{0.5}{\second}, the value of the P-load is increased by $\Delta P_{li}^\ast$ (see Table~\ref{tab:parameters1}). Figure \ref{fig:sen_a} illustrates the time evolution of the system states, showing that the voltages are regulated towards the corresponding references (see Objective~\ref{obj:Volt_regulation}), independently from the load parameters.
Moreover, in order to show the benefits of the passivity property established in Subsection~\ref{subsec:passivity} with respect to the shifted and Krasovskii's passivity properties discussed in Subsection~\ref{subsec:exist_passivity_properties}, we report in Table \ref{tab:parameters3} the values of the equivalent conductance evaluated at the voltage reference, i.e., $G_{\mathrm{B}}(V^*)=G_{\mathrm{K}}(V^*)={Z_l^{*-1}-[P_l^*][V^*]^{-2}}$. Specifically, we observe that before changing the load parameters, the equivalent conductance is positive in each node, implying that the condition $V_i^*\geq\sqrt{P_{li}^*Z_{li}^*}$ is satisfied for all $i=1,\dots,4$  (see Remark~\ref{rm:restrictive_conditions}). However, after changing the load parameters, the equivalent conductance becomes negative and, therefore, $\mathcal{X}_{\mathrm{B}}$ and $\mathcal{X}_{\mathrm{K}}$ defined in \eqref{set:Bregman} and \eqref{set:Krasovskii} do not contain the steady-state solution corresponding to the desired voltage reference $V^\ast$. Conversely, according to the theory developed in the previous sections, the proposed passivity-based controller \eqref{eq:stab_cont} does not require any restriction on the system parameters and is robust with respect to load uncertainties.

%======================================== 

\begin{figure}[t]
	\centering
	\includegraphics[trim=0cm 0cm 0cm 0cm, clip=true, width=\columnwidth]{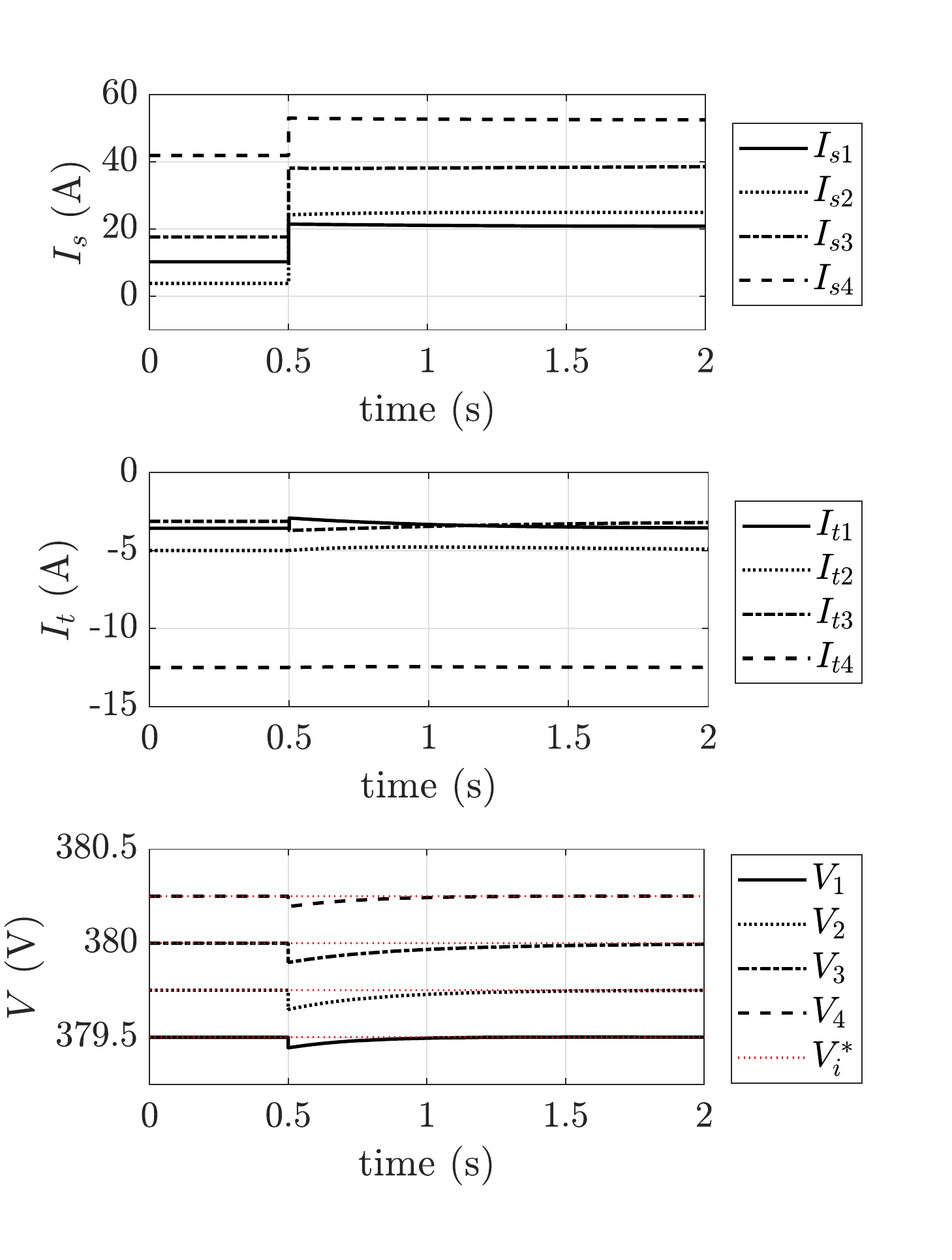}
	\caption{Scenario 2. From the top: time evolution of the generated currents, currents exchanged through the lines, and voltages together with the corresponding references.}
	\label{fig:sen_b}
	\vspace{.2cm}
\end{figure}

{\bf{Scenario  2}} (P-loads). In this second scenario we consider that all the loads of the network are P-loads. 
Differently from the existing passivity properties, which require (for sufficiency) the presence of Z-loads, 
Figure~\ref{fig:sen_b}, illustrates the time evolution of the system states, showing that the voltages are regulated towards the corresponding references (see Objective~\ref{obj:Volt_regulation}), independently from the load parameters. Indeed, as explained in Remark~\ref{rem:controller}, the advantage of having established a passivity property with the output port-variable equal to
the first time derivative of the voltage is the possibility to inject extra damping into the system, counteracting the effects of P-loads independently from the presence or absence of Z-loads.

{If we apply for instance the controller proposed in~\cite[Subsection V.A]{1323174}, the closed-loop system becomes unstable when the value of the power absorbed by the P-loads is sufficiently large.}
  
\begin{table}[t]
	\caption{Network Parameters}
	\centering
	{\begin{tabular}{lc | cccc}			
			Node								&	&1		&2	&3	&4\\			
			\hline
			$R_{si}$	&(\si{\milli\ohm})	&\num{10}		&\num{15}		&\num{25}		&\num{20}\\
			$L_{si}$	&(\si{\milli\henry})	&\num{1.8}		&\num{2.0}		&\num{3.0}		&\num{2.2}\\
			$C_{si}$	&(\si{\milli\farad})	&\num{2.2}		&\num{1.9}		&\num{2.5}		&\num{1.7}\\
			$V_i^{*}$ &(\si{\volt})		&\num{379.50}		&\num{379.75}		&\num{380.00}		&\num{380.25}\\
			$P_{li}^{*}$ &(\si{\kilo\watt})		&\num{10}		&\num{2}		&\num{6}		&\num{10}\\
			$Z_{li}^{*-1}$ &(\si{\siemens})	&\num{0.08}		&\num{0.04}		&\num{0.05}		&\num{0.07}\\
			$I_{li}^*$ &(\si{\ampere})	&\num{10}		&\num{15}		&\num{10}		&\num{15}\\
			$\Delta P_{li}^\ast$ &(\si{\kilo\watt})	&\num{4}		&\num{8}		&\num{8}		&\num{4}\\
			$\Pi_i$ &(\si{\kilo\watt})		&\num{25}		&\num{25}		&\num{25}		&\num{25}\\
	\end{tabular}}
	\label{tab:parameters1}
	\vspace{0.2cm}
	\end{table}
	\vspace{0.2cm}
	
	\begin{table}[t]
	\caption{Line Parameters}
	\centering
	{\begin{tabular}{lc | cccc}			
			Line								&	&1&2	&3		&4\\					
			\hline
			$R_{tk}$	&(\si{\milli\ohm})	&\num{70}		&\num{50}		&\num{80}		&\num{60}\\
			$L_{tk}$	&(\si{\micro\henry})	&\num{2.1}		&\num{2.3}		&\num{2.0}		&\num{1.8}\\
	\end{tabular}}
	\label{tab:parameters2}
	\vspace{.2cm}
\end{table}
\begin{table}[t]
		\caption{Scenario 1: equivalent conductance ${Z_{li}^{*-1}- P_{li}^* {V^*_i}^{-2}}$ (\si{\siemens})}
	\centering
	{\begin{tabular}{l| cccc}			
			Node								&1		&2	&3	&4\\			
			\hline
			for \num{0} $\leq t <$ \num{0.5} \si{\second}	 &\num{0.011}		&\num{0.026}		&\num{0.008}		&\num{0.001}\\
			for \num{0.5} $\leq t \leq$ \num{2} \si{\second} &\num{-0.017}		&\num{-0.029}		&\num{-0.047}		&\num{-0.027}
	\end{tabular}}
	\label{tab:parameters3}
	\vspace{0.2cm}
	\end{table}
	
%
%
%
%
%======================================== End

\section{Conclusions and future works}\label{sec:con}
In this paper we have addressed the notorious instability issue related to the presence of unknown constant power loads in DC networks. We have developed indeed a novel decentralized voltage control scheme that is robust with respect to the uncertainty affecting the loads parameters.
More precisely, inspired by the theory developed by Brayton and Moser, we have proposed a novel passifying input and a storage function that lead, under a very mild assumption (which is generally reasonable for most DC networks), to the establishment of a passivity property where the output port-variable is equal to the first time derivative of the voltage. 
Moreover, differently from the existing results in the literature, where restrictive (sufficient) conditions on the load parameters and voltage reference are assumed to be satisfied, the considered DC network in closed-loop with the proposed passifying input is passive for all the trajectories evolving in the subspace of the state-space where the voltage is positive. 

Interesting future research includes the extension of the proposed approach to DC networks with boost converters and the design of distributed
controllers aimed at guaranteeing current or power sharing among the nodes.

%
%
%
%
%======================================== Biblio

\balance
\bibliographystyle{IEEEtran}
\bibliography{autoref2}
 \end{document}